\documentclass{jfm}
\usepackage{rotating}
\ifCUPmtlplainloaded \else
  \checkfont{eurm10}
  \iffontfound
    \IfFileExists{upmath.sty}
      {\typeout{^^JFound AMS Euler Roman fonts on the system,
                   using the 'upmath' package.^^J}%
       \usepackage{upmath}}
      {\typeout{^^JFound AMS Euler Roman fonts on the system, but you
                   don't seem to have the}%
       \typeout{'upmath' package installed. jfm.cls can take advantage
                 of these fonts,^^Jif you use 'upmath' package.^^J}%
       \providecommand\upi{\pi}%
       \providecommand\upartial{\partial}%
      }
  \else
    \providecommand\upi{\pi}%
    \providecommand\upartial{\partial}%
  \fi
\fi

\ifCUPmtlplainloaded \else
  \IfFileExists{amsbsy.sty}
    {\typeout{^^JFound the 'amsbsy' package on the system, using it.^^J}%
     \usepackage{amsbsy}}
    {\providecommand\boldsymbol[1]{\mbox{\boldmath $##1$}}}
\fi
\usepackage{graphicx}

\def\vec#1{{\boldsymbol #1}}

\newcommand\etal{\mbox{\textit{et al. }}}

\newcommand{\rot}{{\rm curl}\mathop{}}
\newcommand{\const}{{\rm const}}

\title[The flow gradients in the vicinity of a shock wave]{The flow gradients in the vicinity of a shock wave for a thermodynamically imperfect gas}
\author[V.~N.~Uskov and P.~S.~Mostovykh]{\fbox{V\ls L\ls A\ls D\ls I\ls M\ls I\ls R\ns N.\ns U\ls S\ls K\ls O\ls V}\ns \and \ns P\ls A\ls V\ls E\ls L\ns S.\ns M\ls O\ls S\ls T\ls O\ls V\ls Y\ls K\ls H \thanks{Email address for correspondence: mostovykh@gmail.com}}
\affiliation{St.--Petersburg State University, St.--Petersburg, Russia}

\begin{document}

\maketitle

\begin{abstract}
Supersonic vortex plane and axisymmetric flows of non--viscous non--heatconductive gas with arbitrary thermodynamic properties in the vicinity of a steady shock wave are studied. The differential equations describing the gas flow exterior to the discontinuity surface and the dynamic compatibility conditions at this discontinuity are used. The gas flow nonuniformity in the shock vicinity is described by the space derivatives of the gasdynamic parameters at a point on the shock surface. The parameters are the gas pressure, density, velocity vector. The derivatives with respect to the directions of the streamline and normal to it, and of the shock surface and normal to it are considered. Space derivatives of all gasdynamic parameters are expressed through the flow nonisobaric factor along the streamline, the streamline curvature, and the flow vorticity and non--isoenthalpy factors. An algorithm for these factors of the gas flow downstream a shock wave determination is developed. Examples of these factors calculation for imperfect oxygen and thermodynamically perfect gas are presented. The influence coefficients of the upstream flow factors on the downstream flow factors are calculated. As an illustration for flows with upstream Mach number 5 it is shown that the flow vorticity factor is the most influenced by the thermodynamical gas properties. The gas flow in the vicinity of the shock is described by the isolines of gasdynamic parameters. Uniform plane and axisymmetric flows on different distances from the axis of symmetry are examined; the isobars, isopycnics, isotachs and isoclines are used to characterize the downstream flow behind a curved shock in an imperfect gas.
\end{abstract}

\section{Introduction}

Stationary shock waves are formed and their interference is observed in case a stationary supersonic gas flow flows around rigid bodies of complex geometry. The thin spatial zones of shock waves can be modelled as surfaces of gasdynamic discontinuities (GDD). The gas flow exterior to the GDD is assumed non--viscous,  non--heatconductive and thermodynamically equilibrium. The problem of the local flow description in the vicinity of the GDD includes calculation of the gas flow parameters and the first space partial derivatives of parameters downstream it. In (Uskov 1983), see also (Adrianov, Starykh \& Uskov 1995) and (Uskov \& Mostovykh 2010), the former problem is called {\it the zero--order problem}, and the latter~--- {\it the first--order problem}. The first--order problem can be set not only for discontinuities but also for discontinuous characteristics. They are infinitively weak shock waves; gasdynamic parameters remain continuous on their surfaces, whereas its first derivatives are discontinuous.

\looseness=-1 In order to solve the zero--order problem the relations between the gasdynamic parameters on the sides of the GDD are established~--- the {\it dynamic compatibility conditions} (DCC) on it. The DCC on shock waves were first obtained by Rankine (1870). In order to solve the first--order problem, gasdynamic parameters of the flows both upstream and downstream the GDD should be known from the zero--order problem solution. The relations between the first space partial derivatives of gasdynamic parameters on both sides of the discontinuity are called {\it differential dynamic compatibility conditions} (DDCC).

In most studies, beginning from the paper of Rankine (1870), the zero--order problem was solved for a thermodynamically perfect gas. The thermally perfect gas model was also often used. In this model the thermal Clapeyron equation of state is assumed to be satisfied, and the gas specific heats are supposed depending on the gas temperature. In Law (1970) the zero--order problem is solved in the limits of a thermally perfect gas model for oxygen; the gas was supposed to be in dissociation equilibrium. Ando (1981) calculated shocks in carbon dioxide in various models of thermally perfect gas. In these models the vibrational degrees of freedom were taken into account and the gas dissociation was considered.

The Euler equations give several linear relations between the space gasdynamic parameters derivatives. It is therefore possible to select a set of derivatives through which all the rest derivatives can be expressed. The totality of the selected derivatives will be called the {\it basic flow unevennesses}. The first--order problem solution can be reduced to their calculation downstream the shock.

The first--order problem was considered in the works Thomas (1947) and Brown (1950). Thomas (1947) obtained the DDCC for shock waves in plane steady flows of a perfect gas. The DDCC were formulated in terms of the coordinate derivatives of the gasdynamic parameters. The parameters derivatives downstream the shock and the streamline curvature were obtained for a uniform upstream flow. Brown received these derivatives for a non--uniform upstream flow. He described the upstream flow non--uniformity with the streamline curvature, the ratio of the gas pressure and the velocity vector polar angle differentials along the streamline and the flow vorticity.

The first--order problem for a discontinuous characteristic is solved in (Courant \& Friedrichs 1948) for an arbitrary steady gas flow.

Lin \& Rubinov (1948) considered the first--order problem in terms of gasdynamic parameters derivatives with respect to natural directions (the directions tangential and normal to the streamline). They studied plane and axisymmetric isoenergetic flows of a perfect gas. For irrotational upstream flows they obtained two relations between pressure and velocity vector polar angle derivatives along the streamline on both sides of the shock and its curvature. In particular, for the shock normal to the upstream flow at some point, they showed the possibility of infinite curvature. Eckert (1975) examined rotational upstream flows. He determined the pressure and velocity polar angle derivatives along the streamline downstream the shock through similar derivatives in the upstream flow, its vorticity and the shock curvature. Eckert also studied parameter gradients in the vicinity of discontinuous characteristics and rarefaction waves.

Truesdell (1952) analysed imperfect gas flows in the vicinity of shock waves. He calculated the vorticity downstream a curved shock in a uniform upstream flow.

D'yakov (1957) obtained the DDCC in an imperfect gas stream. He used them to study the interaction of a shock wave with a discontinuous characteristic. Rusanov (1973) formulated the DDCC in three--dimensional steady flows of an imperfect gas. He used the local Cartesian coordinate system associated with a point on the shock surface. For a uniform upstream flow he obtained derivatives downstream the shock.

M\"older (1979) considered isoenergetic flows of a perfect gas. He introduced a set of three basic unevennesses in terms of derivatives with respect to natural directions. Derivatives of all parameters and their isolines inclination angles are uniquely determined by them. Relations between the basic unevennesses of the flows on both sides of the shock depending on its curvature are obtained. M\"older (2012) derived equations for the influence coefficients of the upstream flow unevennesses and the shock curvatures on the downstream flow unevennesses. He calculated their dependences on the shock inclination angle for the Mach number $M=3$. The conditions for the Thomas and the Crocco points in a uniform upstream axisymmetric flow are obtained. In (M\"older, Timofeev \& Emanuel 2011; M\"older 2012) the flow in the vicinity of a sonic point on the shock in uniform planar and axisymmetric upstream flows are studied.

In (Uskov 1983) and (Adrianov, Starykh \& Uskov 1995) for plane and axisymmetric flows of a perfect gas a set of three basic flow unevennesses is introduced. They are expressed in terms of the derivatives of static and total pressure and velocity polar angle with respect to natural directions. The analytical formulae for the influence coefficients of the upstream flow unevennesses and the shock curvatures on the downstream flow unevennesses are given. The obtained relations are used for solving the first--order problem for several shock--wave structures.

\looseness=-1 Emanuel \& Liu (1988) received DDCC for shock wave propagating in a nonuniform flow of a single--phase gas in thermodynamic equilibrium. The general theory for the derivatives on the downstream side of a curved shock is developed. The problem is solved in curvilinear coordinates associated with the shock surface. Hornung (1998) examined steady flows of a perfect gas, capable of chemical reactions, in the same coordinates. The rate of energy deposition in the chemical reactions in the flow downstream the shock becomes a significant factor in the problem apart from the shock wave curvature. For a plane uniform upstream flow the influence coefficients of the rate of energy deposition and curvature on the derivatives of gasdynamic 
parameters downstream the shock are obtained.

In this paper the first--order problem is solved for a shock wave; plane and axisymmetric gas flows are considered. The gas on both sides of the GDD satisfies the thermodynamic equation of its state; this equation is assumed to be arbitrary. The gas pressure, density, velocity vector magnitude and its polar angle are chosen as the flow parameters. The problem is solved in terms of derivatives with respect to natural directions.

A standard procedure for the DDCC formulation is realized in \S~2. In \S~3 four basic flow unevennesses are introduced: the flow nonisobaric factor along the streamline $N_1$, the streamline curvature $N_2$, the flow vorticity factor $N_3$ and the flow nonisoenthalpy factor $N_7$. The unevennesses $N_1$, $N_2$, $N_3$ were proposed in (Uskov, 1983) for thermodynamically perfect gas flows; the unevenness $N_7$
is introduced in the present paper. Compared with the papers of M\"older (1979, 2012) the number of unevennesses is increased by one, because the gas flow is not assumed isoenergetic. Three unevennesses of M\"older are expressed through all four unevennesses of the present study.

The influence coefficients of the basic unevennesses of the upstream flow and the shock curvatures on the basic unevennesses of the downstream flow are studied in \S~4. For the case of an overexpanded jet outflow from a nozzle, the shock coming down from the nozzle edge and the jet boundary curvatures are obtained.

In \S~5 the gas flow in the vicinity of the shock, following the ideas of (M\"older 1979), is described by the isolines of the gasdynamic parameters. The behavior of the isolines in the vicinity of the sonic point on the shock surface is investigated. (M\"older \etal 2011) introduced three types of flow in the vicinity of the sonic point; in case of an arbitrary shape of the shock surface four types of flow in this vicinity are distinguished.

\section{Differential dynamic compatibility conditions on shock waves}\nopagebreak

In this section the DDCC on shock waves in arbitrary plane and axisymmetric flows for gases with arbitrary thermodynamic equation of state are obtained. The research methods used to describe shocks in plane and axisymmetric flows are similar. Additional comments on the axisymmetric flow are given in parentheses.

The gas flow at each point is described by its pressure $p$, density $\rho$, temperature $T$, velocity vector $\vec V$. The gas thermodynamic properties are given by the canonical equation of gas state~--- the dependence of its enthalpy $h$ on the pressure $p$ and entropy $s$.

\begin{sidewaysfigure}[htb!]
\vspace{13cm}
\begin{center}
\mbox{\hspace{-0.75cm}\includegraphics[height=6.7cm]{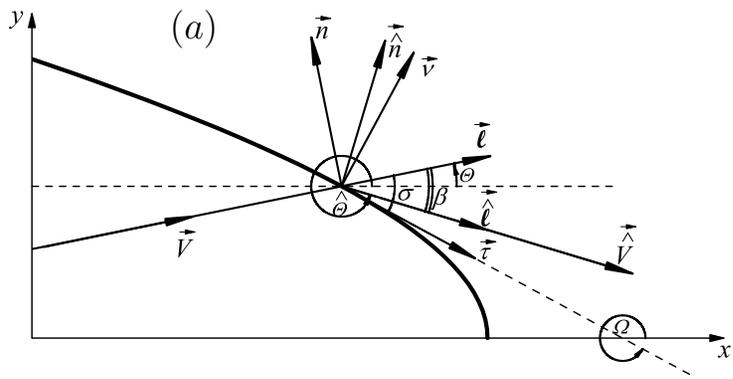} \hspace{-1.1cm}\includegraphics[height=6.7cm]{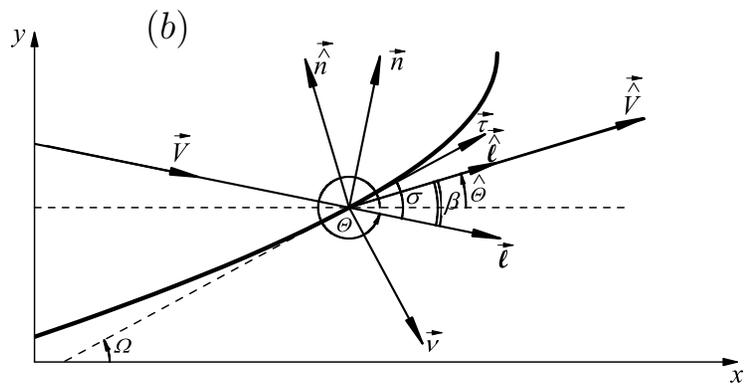}\hspace{-0.85cm}}\\ \mbox{\hspace{-0.95cm}\includegraphics[height=6.7cm]{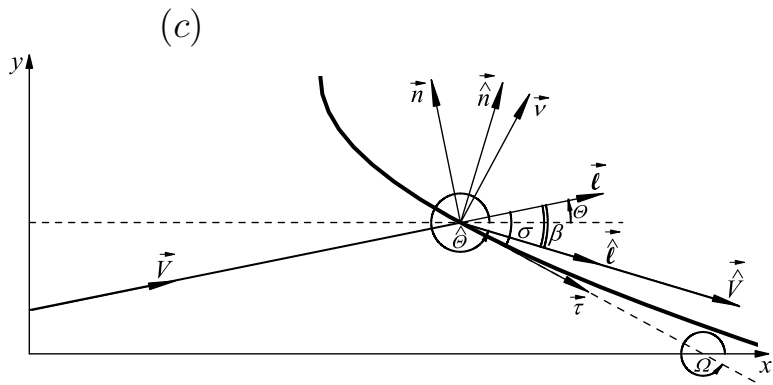} \hspace{-1.1cm}\raisebox{6mm}{\includegraphics[height=6.7cm]{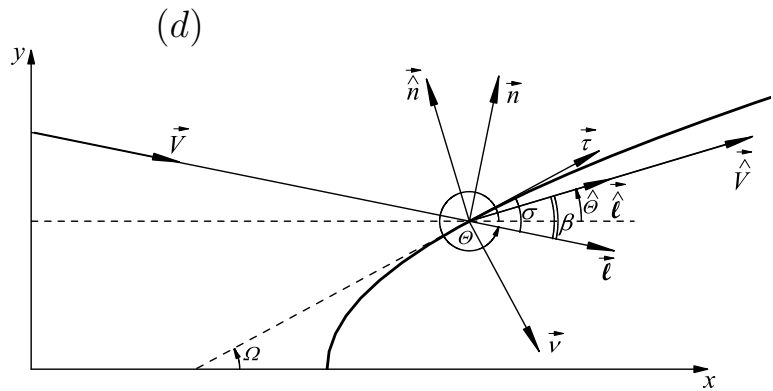}}\hspace{-0.85cm}}\\
\begin{picture}(0,0)
\put(-240,375){\Large $(a)$}
\put(60,375){\Large $(b)$}
\put(-240,185){\Large $(c)$}
\put(60,185){\Large $(d)$}
\end{picture}

\vspace{-1.0cm}

\caption{Flow pictures in the flow plane (meridional half--plane) in the vicinity of an arbitrary point on the shock wave surface.} \label{F1}

\end{center}
\end{sidewaysfigure}

The flow pictures for different geometries of the shock wave surface are shown in figure~\ref{F1}. The flow plane (meridional half--plane) sections are depicted. The shock surfaces on figures~\ref{F1}$(a)$ and~\ref{F1}$(b)$ are concave towards the upstream flow, whereas the shocks on figures~\ref{F1}$(c)$ and~\ref{F1}$(d)$ are convex. For plane gas flows the cases shown on figures~\ref{F1}$(a)$ and~\ref{F1}$(b)$,~\ref{F1}$(c)$ and~\ref{F1}$(d)$ are the same and differ only in the choice of the coordinates system. In axisymmetric flows figures~\ref{F1}$(a)$ and~\ref{F1}$(c)$ correspond to shocks incoming to the axis of symmetry, and figures~\ref{F1}$(b)$ and~\ref{F1}$(d)$~--- to shocks outgoing from the axis. All the notations used in figure~\ref{F1} are explained further on.

Let us introduce three orthogonal unit vectors (orthonormal basis) $\vec \nu$, $\vec \tau$, $\vec b$ (figure~\ref{F1}) at each point of the shock wave. The normal to the shock surface vector $\vec \nu$ is directed to the region downstream the shock wave, the tangential to the shock surface vector $\vec \tau$ belongs to the flow plane (meridional half--plane) and forms an acute angle with the velocity vector $\vec V$, the vector $\vec b$ complements the two vectors to the right--hand system. The gas flow velocity $\vec V$ projections on the directions $\vec \nu$, $\vec \tau$, $\vec b$ are denoted $u_\nu$, $u_\tau$, $u_b$; in this case the following is valid:\pagebreak[0]
\begin{equation}
\displaystyle u_\nu\equiv\left (\vec{V}, \, \vec \nu \right)>0; \quad u_\tau\equiv \left (\vec{V}, \, \vec \tau \right)\geq 0; \quad u_b\equiv \left (\vec{V}, \, \vec b \right)=0; \quad \vec V=u_\nu\vec \nu+u_\tau\vec \tau.\label{1.IV}
\end{equation}

The gasdynamic parameters are discontinuous on the shock wave surface; their derivatives are not defined. Local DCC on a shock wave for an arbitrary point on its surface in terms of the velocity projections have the form (Chernyi 1994):\pagebreak[0]
\begin{equation}
\left \{ \begin{array}{l}
\displaystyle \rho u_\nu=\widehat \rho \widehat u_\nu,\\[0.3cm]
u_\tau=\widehat u_\tau,\\[0.3cm]
p+\rho u_\nu^2=\widehat p+\widehat \rho \widehat u_\nu^2,\\[0.3cm]
\displaystyle h(p, \, \rho)+\frac{V^2}{2}=\widehat h(\widehat{\vphantom{h}p}, \, \widehat{\vphantom{h}\rho})+\frac{\widehat V^2}{2}.
\end{array}\right.\label{1.2}
\end{equation}
Here the parameters marked with $\widehat {\phantom {V}}$ correspond to the gas state downstream the shock, the parameters without it~--- to the gas state upstream it. The gas thermodynamic parameters are interdependent due to the canonical equation of gas state. In particular, the gas enthalpy can be considered as a function of its pressure and density (see Appendix~\ref{App}). Later on the designation $\widehat h$ implies that the enthalpy is calculated for the arguments $\widehat p$, $\widehat \rho$. The flow parameters downstream the shock satisfying the system (\ref{1.2}) are calculated in the zero--order problem. This problem is supposed already solved in the present paper.

Let us introduce the Cartesian coordinates system $x$, $y$ in the flow plane (meridional half--plane). The gas flow velocity vector $\vec V$ belongs to this plane; it can be set by its magnitude $V$ and the polar angle $\varTheta$, which is measured from the $x$ axis to the velocity direction. The positive direction of the polar angles is taken from the direction of the $x$ axis to the $y$ axis. In case of a plane flow the choice of the coordinates system is arbitrary, in case of an axisymmetric flow $x$ is the axis of symmetry, the half--axis $y\geq 0$ is the radius. In the papers of different authors different mutual orientation of the $x$ and $y$ axes was used (right or left). In this paper both cases are included into consideration.

\clearpage The unit vectors of the axes $x$ and $y$ and the unit vectors $\vec \nu$, $\vec \tau$ satisfy the equality $\vec{e}_x \times \vec{e}_y=\chi\vec \nu \times \vec \tau$, the parameter $\chi=\pm 1$ depends on the relative orientation of the $x$, $y$ axes and the vectors $\vec \nu$, $\vec \tau$. The parameter $\chi$ was introduced in (Uskov 1983) as the shock direction index relative to the upstream flow. The index $\chi$ is defined as follows: $\chi=+1$, if the rotation of the gas velocity $\widehat {\vec V}$ downstream the shock towards its surface on the smallest angle (i.e. rotation of $\widehat {\vec V}$ to $\vec \tau$) coincides with the direction of rotation of the $x$ axis to the $y$ axis. Since the $\widehat {\vec V}$ direction lies between the directions $\vec \nu$ and $\vec \tau$, these definitions of the parameter $\chi$ coincide.

Let us denote the shock wave inclination angle $\sigma$ and the flow deflection angle $\beta$ (figure~\ref{F1}): $\sigma$ is the angle between the direction of $\vec V$ upstream the shock and $\vec \tau$, such that $\displaystyle \sigma \in \left(0; \, \upi/2\right]$; $\beta$ is the angle between $\vec V$ and $\widehat {\vec V}$, \mbox{$\beta \in [0; \, \sigma)$}, then the angle between the velocity $\widehat{\vec V}$ direction and the shock surface is $(\sigma-\beta)$. Note that the angles $\sigma$ and $\beta$ are counted off the direction of $\vec \nu$ towards the direction of $\vec \tau$ (this turn coincides with the direction of the $x$ axis towards the $y$ axis in case $\chi=+1$, and is opposite to it in case $\chi=-1$). Using angles $\sigma$ and $\beta$ let us relate the velocities magnitudes $V$ and $\widehat V$ on the sides of the shock with their projections on the directions $\vec \nu$ and $\vec \tau$ and rewrite (\ref{1.IV}), using the second equation (\ref{1.2}), in the form: \pagebreak[0]
\begin{equation}
u_\nu=V \sin \sigma; \quad \widehat {u}_\nu=\widehat V \sin (\sigma-\beta); \quad u_\tau=V \cos \sigma=\widehat V \cos (\sigma-\beta)=\widehat {u}_\tau. \label{1.IL}
\end{equation}

A shock wave can be specified by four parameters of the upstream flow, the shock intensity $J$, and the shock direction index $\chi$ relative to the upstream flow. The gas pressure $p$, temperature $T$, velocity polar angle $\varTheta$, and Mach number $M$ are chosen as parameters. The shock intensity $J\equiv \widehat p/p$ is equal to the ratio of the gas static pressures on the sides of the shock. The zero--order problem is confined to determination of the flow downstream the shock parameters: pressure $\widehat p$, density $\widehat \rho$, velocity polar angle $\widehat \varTheta$ and magnitude $\widehat V$, and the discontinuity inclination angle $\sigma$ and the flow deflection angle $\beta$. Since the shock intensity $J$ and its inclination angle $\sigma$ for the given incoming flow parameters are interdependent (for instance, Mostovykh \& Uskov 2011), the shock can be specified by any of these values; this fact is used later on. Note that in the zero--order problem the input parameters $\varTheta$ and $\chi$ influence only on the angle $\widehat \varTheta$ value:
\begin{equation}
\widehat \varTheta=\varTheta+\chi\beta. \label{hTheta}
\end{equation}
In thermodynamically perfect gas the angles $\sigma$, $\beta$ and $\widehat \varTheta$ do not depend on the pressure $p$ and temperature $T$; the intensity $J$ and the angle $\sigma$ are related by 
$$
J=\frac{2 \gamma}{\gamma+1} M^2\sin^2\sigma-\frac{\gamma-1}{\gamma+1},
$$
here $\gamma$ is the gas specific heats at constant pressure and constant volume ratio.

In order to describe the shock surface, Uskov in (Adrianov \etal 1995) introduced the polar angle $\varOmega$ of the tangential direction vector $\vec \tau$ (figure~\ref{F1}). The angle $\varOmega$ is measured from the $x$ axis towards $\vec \tau$; the angle $\varOmega>0$ in case the rotation takes place in the positive direction. The rotation from the $x$ axis towards the direction $\vec \tau$ can be obtained as a combination of the rotation from the $x$ axis to the direction $\vec V$ on the velocity vector polar angle $\varTheta$ and rotation from $\vec V$ to $\vec \tau$ on the angle $\chi \sigma$, therefore,\pagebreak[0]
\begin{equation}
\varOmega=\varTheta+\chi \sigma. \label{1.XI}
\end{equation}
The polar angle of the direction $\vec \nu$ is $\displaystyle \varOmega-\chi \upi/2$, hence,\pagebreak[0]
\begin{equation}
\vec \nu=\cos \left(\varOmega-\chi \upi/2\right) \vec{e}_x+\sin \left(\varOmega-\chi \upi/2\right) \vec{e}_y.\label{1.XXXVI}
\end{equation}

The derivatives of the normal unit vector $\vec \nu$ with respect to the tangent directions $\vec \tau$ and $\vec b$ define the shock wave curvatures in two mutually perpendicular normal sections: in the flow plane (meridional half--plane) and in the plane perpendicular to it. M\"older (1979) described the shock surface with these curvatures, and designated them $S_a$ and $S_b$, respectively. The curvatures are supposed positive if the shock curve is concave towards the upstream flow in the relevant normal cross--section, and are supposed negative otherwise. Under this choice of signs the equalities
\begin{equation}
\frac {\upartial \vec \nu} {\upartial \tau}=S_a \vec \tau, \qquad \frac {\upartial \vec \tau} {\upartial \tau}=-S_a \vec \nu; \qquad \qquad \frac {\upartial \vec \nu} {\upartial b}=S_b \vec b, \qquad \frac {\upartial \vec b} {\upartial b}=-S_b \vec \nu \label{1.XXIV}
\end{equation}
for the unit vectors derivatives with respect to the tangential directions hold. In the plane flow the curvature $S_b$ turns to zero.

In (Rusanov 1973) the discontinuity curvatures are introduced in the same way as in (M\"older 1979), but their signs are the opposite ones.

In (Uskov 1983) and (Adrianov \etal 1995) the shock wave surface is described by the curvatures $N_4$ and $N_5$. The curvature $N_5$ in the flow plane (meridional half--plane) is defined by the formula $N_5 \equiv \upartial \varOmega / \upartial \tau$. The curvature $N_4$ in plane flows is assumed to be zero, in axisymmetric flows it is determined by the shock radius in a plane perpendicular to the axis of symmetry. Precisely, $\displaystyle N_4 \equiv \delta/y$, here for plane flows $\delta=0$, for axisymmetric flows $\delta=1$, so that $N_4 \geq 0$.

Let us establish a relation between the curvatures $S_a$, $S_b$ and $N_5$, $N_4$. Differentiating (\ref{1.XXXVI}) with respect to $\vec \tau$, we get \pagebreak[0]
$$
\frac{\upartial \vec \nu}{\upartial \tau}=\left (\chi \cos \varOmega \vec {e}_x + \chi \sin \varOmega \vec {e}_y \right) N_5=\chi N_5 \vec \tau,
$$
thus, \pagebreak[0]
\begin{equation}
S_a=\chi N_5. \label{1.6p}
\end{equation}
According to the Meusnier theorem (Smirnov 1964), $N_4$ and $S_b$ differ by a factor equal to the cosine of the angle between the normal to the shock surface vector $\vec \nu$ and the unit vector $\vec{e}_y$ of the radius. Thus,
$$
S_b=N_4 \cos \left (\varOmega-\chi \frac{\upi}{2}-\frac{\upi}{2} \right)=- \chi N_4 \cos \varOmega.
$$

In this paper, the curvatures $S_a$ and $N_4$ are used to describe the shock surface. Selecting $N_4$ instead of $S_b$ avoids the uncertainty of the form $0/0$ in the description of shocks perpendicular to the axis of symmetry ($\varOmega=\pm \upi/2$).

\begin{table}
\begin{center}
\begin{tabular}{lcccc}
&Figure~\ref{F1}$(a)$&Figure~\ref{F1}$(b)$&Figure~\ref{F1}$(c)$&Figure~\ref{F1}$(d)$\\
$\chi$&$-1$&$+1$&$-1$&$+1$\\
$S_a$&$>0$&$>0$&$<0$&$<0$\\
$S_b$&$\geq 0$&$\leq 0$&$\geq 0$&$\leq 0$\\
$N_5$&$<0$&$>0$&$>0$&$<0$\\
$\varOmega$&$\left[3\upi/2;\, 2\upi\right]$&$\left[0;\, \upi/2\right]$&$\left[3\upi/2;\, 2\upi\right]$&$\left[0;\, \upi/2\right]$
\end{tabular}
\end{center}
\caption{The parameters corresponding to the different geometries of the shock surface}\label{tabl1}
\end{table}

Table~\ref{tabl1} shows the values of the shock direction index $\chi$ relative to the incident flow, the signs of the curvatures $S_a$, $S_b$ and $N_5$, the intervals of the polar angle $\varOmega$ change, corresponding to the shock geometries shown in figure~\ref{F1}. Let us note that if in an axisymmetric flow the gas flows in the positive direction of the $x$ axis ($\left|\varTheta\right|<\upi/2$), the value $\chi=-1$ corresponds to a shock, incoming to the axis of symmetry, i.e. $S_b>0$, and the value $\chi=+1$~--- to a shock, outcoming from the axis, i.e. $S_b<0$. In other words, the sign of the previously introduced direction index $\chi$ is always opposite to the curvature $S_b$ sign.

The relationship between the gasdynamic parameters derivatives along the surface of the shock wave on the two sides of it is established by differentiating relations (\ref{1.2}) along its surface, i.e. in the directions $\vec \tau$ and $\vec b$:\pagebreak[0]
\begin{equation}
\left \{ \begin{array}{l} \displaystyle \frac{\upartial \rho}{\upartial \tau} u_\nu + \rho \frac{\upartial u_\nu}{\upartial \tau}=\frac{\upartial \widehat \rho}{\upartial \tau} \widehat{u}_\nu + \widehat{\rho} \frac{\upartial \widehat{u}_\nu}{\upartial \tau},\\[0.5cm]
\displaystyle \frac{\upartial u_\tau}{\upartial \tau}=\frac{\upartial \widehat{u}_\tau}{\upartial \tau},\\[0.5cm]
\displaystyle \frac{\upartial p}{\upartial \tau} + \frac{\upartial \rho}{\upartial \tau} u_\nu^2 +2 \rho u_\nu \frac{\upartial u_\nu}{\upartial \tau}=\frac{\upartial \widehat {p}}{\upartial \tau} + \frac{\upartial \widehat {\rho}}{\upartial \tau} \widehat{u}_\nu^2 +2\widehat{\rho} \widehat{u}_\nu \frac{\upartial \widehat{u}_\nu}{\upartial \tau},\\[0.5cm]
\displaystyle h_p \frac{\upartial p}{\upartial \tau}+h_\rho \frac{\upartial \rho}{\upartial \tau} + V \frac{\upartial V}{\upartial \tau}=\widehat{h}_p \frac{\upartial \widehat{p}}{\upartial \tau}+\widehat{h}_\rho \frac{\upartial \widehat{\rho}}{\upartial \tau}+ \widehat V \frac{\upartial \widehat V}{\upartial \tau},
\end{array} \right. \label{1.VI}
\end{equation}
here $\displaystyle h_p\equiv \upartial h/\upartial p \bigl |_{\rho=\const}$ and $\displaystyle h_\rho \equiv \upartial h/\upartial \rho \bigl |_{p=\const}$. In the plane case, all gasdynamic parameters remain constant in the direction $\vec b$, since $\vec b$ is perpendicular to the flow plane. In the axisymmetric case, the change in the direction $\vec b$ corresponds to a transition to a different meridional half--plane with the same values of all the scalar gasdynamic parameters. Consequently, all derivatives in the direction $\vec b$ are zero; the equations that arise as a result of differentiating (\ref{1.2}) in the direction $\vec b$, hold identically both in plane and axisymmetric cases.

The system (\ref{1.VI}) can be considered as an algebraic system of equations with respect to the gasdynamic parameters derivatives $\upartial \widehat p / \upartial \tau$, $\upartial \widehat \rho / \upartial \tau$, $\upartial \widehat{u}_\nu / \upartial \tau$, $\upartial \widehat{u}_\tau / \upartial \tau$. These derivatives are taken along the tangent to the shock surface at some fixed point on the downstream side of the shock. The unknown $\upartial \widehat V / \upartial \tau$ can be expressed using (\ref{1.IL}):
$$
\widehat V^2=\widehat{u}_\nu^2+\widehat{u}_\tau^2; \qquad \widehat V\frac{\upartial \widehat V}{\upartial \tau}=\widehat{u}_\nu\frac{\upartial \widehat{u}_\nu}{\upartial \tau}+\widehat{u}_\tau\frac{\upartial \widehat{u}_\tau}{\upartial \tau}.
$$
After division by $\widehat V$, we get
$$
\frac{\upartial \widehat V}{\upartial \tau}=\frac{\upartial \widehat{u}_\nu}{\upartial \tau}\sin (\sigma-\beta)+\frac{\upartial \widehat{u}_\tau}{\upartial \tau}\cos (\sigma-\beta).
$$

Let us deduce the DDCC on a shock wave in terms of the derivatives of gas pressure, density, and velocity magnitude and polar angle. The projections of the velocity vector $u_\nu $, $u_\tau$ on the shock surface depend on the velocity $\vec V$ and the basis vectors $\vec \nu$ and $\vec \tau$. Let us express the derivatives of $u_\nu $, $u_\tau$ with respect to the tangential direction $\vec \tau$ through the derivatives of the velocity magnitude $V$, its polar angle $\varTheta$, and the shock curvature. The derivative $\upartial u_\nu/\upartial \tau$ is given by differentiation of (\ref{1.IV}):\pagebreak[0]
\begin{equation}
\frac{\upartial u_\nu}{\upartial \tau}\equiv \frac{\upartial}{\upartial \tau} \left (\vec{V}, \, \vec \nu \right)=\left(\frac{\upartial \vec{V}}{\upartial \tau}, \, \vec \nu \right) +\left (\vec{V}, \, \frac{\upartial \vec \nu}{\upartial \tau} \right). \label{1.Va}
\end{equation}
Let us fix a point $O$ on the shock surface; all the quantities referring to the point $O$ will be marked with the index $O$. The vectors $\vec \nu_O$, $\vec \tau_O$ define an orthonormal basis in the flow plane (meridional half--plane), consequently, $\vec{V}=\left(\vec{V}, \, \vec \nu_O \right)\vec \nu_O+\left(\vec{V}, \, \vec \tau_O \right)\vec \tau_O$ in the vicinity of $O$. For the derivatives in the direction $\vec \tau$ we have\pagebreak[0]
\begin{equation}
\frac{\upartial \vec{V}}{\upartial \tau}=\frac{\upartial \left(\vec{V}, \, \vec \nu_O \right)}{\upartial \tau}\vec \nu_O+\frac{\upartial \left(\vec{V}, \, \vec \tau_O \right)}{\upartial \tau}\vec \tau_O.\label{1.XXII}
\end{equation}
The polar angles of the velocity vector $\vec V$ and the directions $\vec \nu_O$ and $\vec \tau_O$ are equal to $\varTheta$, $\displaystyle \varOmega_O-\chi \upi/2$ and $\varOmega_O$, respectively, hence at any point of the shock we obtain\pagebreak[0]
$$
\frac{\upartial \left(\vec{V}, \, \vec \nu_O \right)}{\upartial \tau}=\frac{\upartial }{\upartial \tau}\left(V \cos \left(\varTheta-\left(\varOmega_O-\chi\frac{\upi}{2}\right)\right)\right)=\chi\frac{\upartial}{\upartial \tau}\left(V \sin(\varOmega_O-\varTheta)\right)=
$$
$$
=\chi\frac{\upartial V}{\upartial \tau}\sin(\varOmega_O-\varTheta)-\chi V \cos(\varOmega_O-\varTheta)\frac{\upartial \varTheta}{\upartial \tau};
$$
$$
\frac{\upartial \left(\vec{V}, \, \vec \tau_O \right)}{\upartial \tau}=\frac{\upartial }{\upartial \tau}\left(V \cos (\varTheta-\varOmega_O)\right)=\frac{\upartial V}{\upartial \tau}\cos(\varOmega_O-\varTheta) +V \sin(\varOmega_O-\varTheta)\frac{\upartial \varTheta}{\upartial \tau}.
$$
Using (\ref{1.XXII}), (\ref{1.XI}) and (\ref{1.IL}), for the derivatives at the point $O$ we get\pagebreak[0]
\begin{equation}
\begin{array}{l}
\displaystyle \left(\frac{\upartial \vec{V}}{\upartial \tau}, \, \vec \nu_O \right)=\frac{\upartial \left(\vec{V}, \, \vec \nu_O \right)}{\upartial \tau}= \sin \sigma_O\frac{\upartial V}{\upartial \tau}-\chi (u_\tau)_O \frac{\upartial \varTheta}{\upartial \tau},
\end{array}\label{1.XXXI}
\end{equation}
\begin{equation}
\begin{array}{l}
\displaystyle \left(\frac{\upartial \vec{V}}{\upartial \tau}, \, \vec \tau_O \right)=\frac{\upartial \left(\vec{V}, \, \vec \tau_O \right)}{\upartial \tau}= \cos \sigma_O\frac{\upartial V}{\upartial \tau}+\chi (u_\nu)_O \frac{\upartial \varTheta}{\upartial \tau}.
\end{array}\label{1.XXX}
\end{equation}

The derivative $\upartial u_\nu/\upartial \tau$, given by (\ref{1.Va}), at the point $O$ using (\ref{1.XXXI}), (\ref{1.XXIV}) and (\ref{1.IV}) reduces to\pagebreak[0]
\begin{equation}
\begin{array}{l}
\displaystyle \frac{\upartial u_\nu}{\upartial \tau}= \sin \sigma_O\frac{\upartial V}{\upartial \tau}-\chi (u_\tau)_O
\frac{\upartial \varTheta}{\upartial \tau}+(u_\tau)_O S_a. \end{array}\label{1.XL}
\end{equation}
Let us consider the second equation (\ref{1.VI}) at the point $O$ on the shock and re--arrange its left side, using (\ref{1.IV}), (\ref{1.XXX}), (\ref{1.XXIV}):\pagebreak[0]
\begin{equation}
\begin{array}{l}
\displaystyle \frac{\upartial u_\tau}{\upartial \tau}=\frac{\upartial \left(\vec V, \, \vec \tau \right)}{\upartial \tau}=\left(\frac{\upartial \vec{V}}{\upartial \tau}, \, \vec \tau_O \right)+\left(\vec V_O, \, \frac{\upartial \vec \tau}{\upartial \tau}\right)=\\[0.5cm]
\displaystyle \qquad \qquad \qquad \qquad \qquad \qquad =\cos \sigma_O\frac{\upartial V}{\upartial \tau}+\chi (u_\nu)_O\frac{\upartial \varTheta}{\upartial \tau}-(u_\nu)_O S_a. \end{array}\label{1.XLI}
\end{equation}
In order to adapt the formulae (\ref{1.XL}), (\ref{1.XLI}) to derivatives downstream the shock wave, it is necessary to add $\widehat {\phantom{V}}$ and to replace $\sigma_O$ with $\sigma_O-\beta_O$. In the following the parameters and their derivatives are calculated at $O$. For simplicity, the index $O$ is below omitted.

Let us rewrite (\ref{1.VI}), using (\ref{1.XL}), (\ref{1.XLI}), and taking $u_\tau=\widehat{u}_\tau$ (second DCC (\ref{1.2})) into account:\pagebreak[0]
\begin{equation}
\left \{ \begin{array}{l}
\displaystyle \frac{\upartial \rho}{\upartial \tau} u_\nu + \rho \sin \sigma\frac{\upartial V}{\upartial \tau}-\chi \rho u_\tau \frac{\upartial \varTheta}{\upartial \tau}+S_a \rho u_\tau=\\[0.5cm]
\qquad \qquad \displaystyle=\frac{\upartial \widehat \rho}{\upartial \tau} \widehat{u}_\nu + \widehat \rho \sin (\sigma-\beta)\frac{\upartial \widehat V}{\upartial \tau}-\chi \widehat \rho u_\tau \frac{\upartial \widehat \varTheta}{\upartial \tau}+S_a \widehat \rho u_\tau,\\[0.5cm]
\displaystyle \cos \sigma\frac{\upartial V}{\upartial \tau}+\chi u_\nu\frac{\upartial \varTheta}{\upartial \tau}-u_\nu S_a=\cos (\sigma-\beta)\frac{\upartial \widehat V}{\upartial \tau}+\chi \widehat {u}_\nu \frac{\upartial \widehat \varTheta}{\upartial \tau}-\widehat{u}_\nu S_a,\\[0.5cm]
\displaystyle \frac{\upartial p}{\upartial \tau} + \frac{\upartial \rho}{\upartial \tau} u_\nu^2 +2 \rho u_\nu \sin \sigma\frac{\upartial V}{\upartial \tau}-2 \chi \rho u_\nu u_\tau \frac{\upartial \varTheta}{\upartial \tau}+\underline{2 S_a \rho u_\nu u_\tau}=\\[0.5cm]
\qquad \displaystyle=\frac{\upartial \widehat p}{\upartial \tau} + \frac{\upartial \widehat \rho}{\upartial \tau} \widehat{u}_\nu^2 +2 \widehat \rho \widehat{u}_\nu \sin (\sigma-\beta)\frac{\upartial \widehat V}{\upartial \tau}-2 \chi \widehat{\rho} \widehat{u}_\nu u_\tau \frac{\upartial \widehat \varTheta}{\upartial \tau}+\underline{2 S_a \widehat \rho \widehat{u}_\nu u_\tau},\\[0.5cm]
\displaystyle h_p \frac{\upartial p}{\upartial \tau}+h_\rho \frac{\upartial \rho}{\upartial \tau}+V\frac{\upartial V}{\upartial \tau}=\widehat{h}_p \frac{\upartial \widehat{p}}{\upartial \tau}+\widehat{h}_\rho \frac{\upartial \widehat \rho}{\upartial \tau}+\widehat V\frac{\upartial \widehat V}{\upartial \tau}.
\end{array}\right.\hspace{-1cm} \label{1.XX}
\end{equation}
The underlined terms can be reduced on the basis of the first equation (\ref{1.2}). The system (\ref{1.XX}) gives the DDCC on a shock wave.

\section{The basic gas flow unevennesses}\nopagebreak

In this section a full set of the basic gas flow unevennesses is developed; the first derivatives of all gasdynamic parameters with respect to an arbitrary direction can be calculated in terms of this set. The relations between this set and the unevennesses, introduced earlier, for special cases of gas flows, in (M\"older 1979) and (Uskov 1983), are established.

The basic gas flow unevennesses should be determined by the gasdynamic parameters variation in space and do not depend on the choice of the coordinates system. In order to define them, let us introduce the unit vectors of natural directions $\vec \ell$ and $\vec n$, directed along the streamlines (co--directional to the gas velocity) and the normal to it, respectively, at each point of the flow. Direction of the normal $\vec n$ is chosen so that the rotation of $\vec \ell$ to $\vec n$ occurs in the positive direction, i.e., that at any point the equality $\vec e_x \times \vec e_y=\vec \ell \times \vec n$ holds. The gasdynamic parameters derivatives with respect to the natural directions are independent of the coordinates system, and are fully determined by the gas flow.

The gas dynamics equations for plane and axisymmetric steady motion of non--viscous non--heatconductive gas in the natural directions $\vec \ell$, $\vec n$ have the form (Hayes \& Probstein 1966; Ginsburg 1966):\pagebreak[0]
$$
\left \{\hspace{-7.3cm}
\vphantom{\begin{array}{l}
\displaystyle \frac Ae\\[0.25cm]
\displaystyle \frac Bf\\[0.25cm]
\displaystyle \frac Cg\\[0.25cm]
\displaystyle \frac cg\\[0.25cm]
\displaystyle \frac Dh
\end{array}} \right.\eqno{\parbox{13.4cm}{
\begin{subeqnarray}
\gdef \thesubequation {\theequation \textit{a}}
\displaystyle &&\rho\frac{\upartial V}{\upartial \ell}+\displaystyle V \frac{\upartial \rho}{\upartial \ell}+\rho V \frac{\upartial \varTheta}{\upartial n}+\frac{\delta}{y}\rho V\frac{\upartial y}{\upartial \ell}=0, \\[0.2cm] \slabel{1.3a}
\gdef \thesubequation {\theequation \textit{b}}
\displaystyle &&\rho V \frac{\upartial V}{\upartial \ell}=\displaystyle -\frac{\upartial p}{\upartial \ell}, \\[0.2cm] \slabel{1.3b}
\gdef \thesubequation {\theequation \textit{c}}
\displaystyle &&\rho V^2 \frac{\upartial \varTheta}{\upartial \ell}=\displaystyle -\frac{\upartial p}{\upartial n}, \\[0.2cm] \slabel{1.3c}
\gdef \thesubequation {\theequation \textit{d}}
\displaystyle &&s_p \frac{\upartial p}{\upartial\ell}+\displaystyle s_\rho \frac{\upartial \rho}{\upartial \ell}=0. \slabel{1.3d} \label{1.3}
\end{subeqnarray}
\returnthesubequation
}}
$$
The dependence of the gas specific entropy on its pressure and density $s(p, \, \rho)$ and its derivatives is determined from the canonical equation of gas state (see appendix~\ref{App}).

All flow parameters at a given point can be expressed algebraically through four parameters $p$, $\rho$, $V$ and $\varTheta$. First derivatives of these parameters can be expressed algebraically in terms of the $p$, $\rho$, $V$ and $\varTheta$ derivatives with respect to two mutually perpendicular directions, for example, $\vec \ell$ and $\vec n$. Consequently, these eight derivatives fully describe the non--uniformity of the flow at a given point. System (\ref{1.3}) gives four equations, linearly relating these derivatives. Thus, at every point, there are four linearly independent flow parameters first derivatives; their number coincides with the number of differential equations describing the gas flow.

In order to characterize the non--uniformity of the gas flow at an arbitrary point it is necessary and sufficient to specify four factors, which are called {\it basic gas flow unevennesses}. Uskov (1983) and M\"older (1979) independently of each other introduced the sets of three basic gas flow unevennesses. Both authors expressed the basic unevennesses in terms of the derivatives with respect to natural directions. They are:\pagebreak[0]
\begin{equation}
\left\{ \begin{array}{l}
\displaystyle N_1\equiv \frac{\upartial \ln p}{\upartial\ell}, \\[0.5cm]
\displaystyle N_2\equiv \frac{\upartial \varTheta}{\upartial\ell}, \\[0.5cm]
\displaystyle N_3\equiv \frac{\upartial \ln p_0}{\upartial n},
\end{array}
\right. \qquad \qquad \qquad \qquad\left\{ \begin{array}{l}
\displaystyle P\equiv \frac{1}{\rho V^2}\frac{\upartial p}{\upartial\ell}, \\[0.5cm]
\displaystyle D\equiv \frac{\upartial \varTheta}{\upartial\ell}, \\[0.5cm]
\displaystyle \varGamma\equiv -\frac{\left(\rot \vec V\right)_\perp}{V}.
\end{array}\right.\label{1.4}
\end{equation}
The vector $\rot \vec V$ is perpendicular to the flow plane, the kinematic vorticity $\left(\rot \vec V \right)_\perp$ is its projection on the direction forming a right--hand system with the pairs $\vec \ell$ and $\vec n$, $\vec e_x$ and $\vec e_y$. On the left are the basic unevennesses of the gas flow, introduced in (Uskov 1983), on the right~--- introduced in (M\"older 1979). Let us explain their physical sense: $N_1$~--- the flow nonisobaric factor along the streamline, $N_2$~--- the streamline curvature, $N_3$~--- the flow vorticity factor; $P$~--- the dimensionless pressure gradient projected on the streamline, $D$~--- the streamline curvature, $\varGamma$~--- the flow vorticity. Note that all of the basic unevennesses have the dimension m$^{-1}$.

In (\ref{1.4}) the gas flow rest pressure $p_0$ is used, which, together with the rest density $\rho_0$ satisfies the system of equations:\pagebreak[0]
\begin{equation}
\left \{ \begin{array}{l}
\displaystyle s(p_0, \, \rho_0)=s(p, \, \rho),\\[0.3cm]
\displaystyle h(p_0, \, \rho_0)=h(p, \, \rho) +\frac{V^2}{2}.
\end{array}\right. \label{1.5}
\end{equation}
In the following, the gas flow rest parameters are denoted with subscript $0$. The subscript $0$ is also used for derivatives of $s$ and $h$ calculated at the values $p_0$, $\rho_0$.

In (M\"older 1979) only flows with constant rest enthalpy in the whole flow field were considered: $h_0=\const$. This is a broad class of flows, among their number, unseparated flows around solid bodies of any shape by a uniform initial flow. However, the condition of constant rest enthalpy breaks in both the wall and the separated boundary layer. Assuming rest enthalpy to be constant, the parameters $p$, $\rho$ and $V$ are algebraically related by the second relation (\ref{1.5}), and three independent equations remain in the system (\ref{1.3}). This reduces the number of independent first derivatives to three.

In (Uskov 1983) and (Adrianov \etal 1995), the problem was solved for a thermodynamically perfect gas. In this case the gas flow dynamic pressure $d=\rho V^2/2$, static pressure $p$ and the velocity vector polar angle $\varTheta$ meet a system of three equations:
$$
\left\{ \hspace{-8.3cm} \vphantom{\begin{array}{l}
\displaystyle \frac Ae\\[0.1cm]
\displaystyle \frac Bf\\[0.1cm]
\displaystyle \frac Cg\\[0.1cm]
\displaystyle \frac cg\\[0.1cm]
\end{array}
} \right.\eqno{\parbox{13.4cm}{
\begin{subeqnarray}
\gdef \thesubequation {\theequation \textit{a}}
\displaystyle &&2\frac{\upartial d}{\upartial \ell}+2d \frac{\upartial \varTheta}{\upartial n}+2\frac{\delta}{y}d\frac{\upartial y}{\upartial \ell}=-\frac{\upartial p}{\upartial \ell}, \\[0.2cm] \slabel{1.30a}
\gdef \thesubequation {\theequation \textit{b}}
\displaystyle &&-\frac{V^2}{2a^2} \frac{\upartial p}{\upartial\ell}+ \frac{\upartial d}{\upartial \ell}=-\frac{\upartial p}{\upartial \ell}, \\[0.2cm] \slabel{1.30b}
\gdef \thesubequation {\theequation \textit{c}}
\displaystyle &&2d \frac{\upartial \varTheta}{\upartial \ell}=-\frac{\upartial p}{\upartial n}, \slabel{1.30c} \label{1.30}
\end{subeqnarray}
\returnthesubequation
}}
$$
here $a^2=-s_\rho/s_p$ is squared the local speed of sound in the gas. The equation (\ref{1.30a}) is a sum of (\ref{1.3a}) multiplyed by $V$ and (\ref{1.3b}); (\ref{1.30b}) is a sum of (\ref{1.3d}) multiplyed by $V^2/(2s_\rho)$ and (\ref{1.3b}); (\ref{1.30c}) is (\ref{1.3c}). The relations (\ref{1.30}) for $d$, $p$, $V$ and $\varTheta$ are valid for a gas with arbitrary properties. For a thermodynamically perfect gas $V^2/(2a^2)=d/(\gamma p)$, and the system (\ref{1.30}) is a closed system of three equations with respect to $d$, $p$ and $\varTheta$. The number of the independent first derivatives of parameters also reduces to three. Let us note that the system (\ref{1.30}) does not allow to determine all flow parameters. In particular, it is impossible to determine $\rho$ and $V$ and, consequently, their derivatives, rest enthalpy $h_0$ and rest density $\rho_0$. Therefore, the assumption of constant rest enthalpy, made in (M\"older 1979), can not be considered within the limits of the system (\ref{1.30}).

In the following the problem is solved for a gas with arbitrary thermodynamic properties. The basic unevennesses $N_1$, $N_2$ and $N_3$ according to (Uskov 1983) and the value\pagebreak[0]
\begin{equation}
N_7\equiv \frac{\rho_0}{p_0}\frac{\upartial h_0}{\upartial n}=\frac{\rho_0}{p_0}\left(h_{p0}\frac{\upartial p_0}{\upartial n} +h_{\rho 0}\frac{\upartial \rho_0}{\upartial n}\right), \label{1.I}
\end{equation}
the flow non--isoenthalpy factor, are used (index 6 for $N$ was previously used in (Uskov \& Mostovykh 2008)). M\"older (1979) assumed $N_7=0$.

The first two pairs of basic unevennesses in (\ref{1.4}) coincide up to a factor that depends only on the gasdynamic parameters at the point (but independent of its derivatives):\pagebreak[0]
$$
P=\frac{p}{\rho V^2}N_1; \qquad D=N_2.
$$
In order to establish a relationship between the vorticity $\varGamma$ and the unevennesses $N_i$\linebreak (\mbox{$i=1, \, 2, \, 3, \, 7$}), let us differentiate (\ref{1.5}) in the direction $\vec n$:\pagebreak[0]
\begin{equation}
\left \{ \begin{array}{l}
\displaystyle s_p \frac{\upartial p}{\upartial n}+s_\rho \frac{\upartial \rho}{\upartial n}=s_{p0} \frac{\upartial p_0}{\upartial n}+s_{\rho 0} \frac{\upartial \rho_0}{\upartial n},\\[0.5cm]
\displaystyle h_p \frac{\upartial p}{\upartial n}+h_\rho \frac{\upartial \rho}{\upartial n}+ V \frac{\upartial V}{\upartial n}=\frac{\upartial h_0}{\upartial n}.
\end{array} \right. \label{1.6}
\end{equation}
The derivative $\displaystyle \upartial p/\upartial n$ is expressed through the unevenness $N_2$ using the third equation (\ref{1.3}). Then from (\ref{1.I}) and (\ref{1.6}) we get\pagebreak[0]
$$
\frac{\upartial \rho_0}{\upartial n}=\frac{p_0}{\rho_0h_{\rho 0}}N_7-\frac{h_{p0}}{h_{\rho 0}}\frac{\upartial p_0}{\upartial n}=\frac{p_0}{\rho_0h_{\rho 0}}N_7-\frac{p_0h_{p0}}{h_{\rho 0}}N_3;
$$
\begin{equation}
\begin{array}{l}
\displaystyle \frac{\upartial \rho}{\upartial n}=\frac{s_{p0}}{s_\rho}\frac{\upartial p_0}{\upartial n}+\frac{s_{\rho 0}}{s_\rho}\frac{\upartial \rho_0}{\upartial n}-\frac{s_p}{s_\rho}\frac{\upartial p}{\upartial n}=\\[0.3cm]
\displaystyle \qquad \qquad \qquad =\frac{p_0}{s_\rho}\left(s_{p0}-\frac{s_{\rho 0}h_{p0}}{h_{\rho 0}}\right)N_3+\frac{p_0s_{\rho 0}}{\rho_0h_{\rho 0}s_\rho}N_7+\frac{s_p\rho V^2}{s_\rho}N_2. \end{array}\label{1.L}
\end{equation}

Expressing the kinematic vorticity $\left( \rot \vec V\right)_\perp$ in the system of natural directions and using (\ref{1.I}), (\ref{1.6}) and (\ref{1.L}), we get for the vorticity $\varGamma$\pagebreak[0]
$$
\varGamma=\frac{\upartial \ln V}{\upartial n}-\frac{\upartial \varTheta}{\upartial\ell}=\frac{1}{V^2}\left(\frac{p_0}{\rho_0}N_7- h_p\frac{\upartial p}{\upartial n}-h_\rho\frac{\upartial \rho}{\upartial
n}\right)-\frac{\upartial \varTheta}{\upartial\ell}=
$$
$$
=\frac{p_0}{\rho_0V^2}\left(1-\frac{h_\rho s_{\rho 0}}{h_{\rho 0}s_\rho}\right)N_7-\frac{p_0h_\rho}{V^2s_\rho}\left(s_{p0}-\frac{s_{\rho 0}h_{p0}}{h_{\rho 0}}\right)N_3+ \left(\rho h_p-\frac{s_p\rho h_\rho}{s_\rho}-1\right)N_2.
$$
Using the thermodynamic identity (\ref{1.III}), we get:\pagebreak[0]
\begin{equation}
\begin{array}{l}
\displaystyle \varGamma=\frac{p_0}{\rho_0V^2}\left(1-\frac{h_\rho s_{\rho 0}}{h_{\rho 0}s_\rho}\right)N_7+\frac{p_0h_\rho}{V^2s_\rho}\cdot \frac{s_{\rho 0}}{\rho_0h_{\rho 0}}N_3=\\[0.5cm]
\displaystyle \qquad \qquad \qquad \qquad \qquad \qquad=\frac{p_0}{\rho_0V^2}N_7+\frac{p_0}{\rho_0V^2}\cdot \frac{h_\rho s_{\rho 0}}{h_{\rho 0}s_\rho}(N_3-N_7).\end{array}\label{1.LII}
\end{equation}
For a thermally perfect gas (i.e., gas satisfying the Clapeyron equation of state $p=\rho RT$), the last formula is simplified:\pagebreak[0]
$$
\varGamma=\frac{p_0}{\rho_0V^2}N_7+\frac{p}{\rho V^2}(N_3-N_7),
$$
and for thermally and calorically perfect gas, we have\pagebreak[0]
$$
\varGamma=\frac{N_3}{\gamma M^2}+\frac{\gamma-1}{2\gamma}N_7.
$$
As already mentioned, in (M\"older 1979) the equality $N_7=0$ is assumed, then $N_3$ and $\varGamma$ are related by $\displaystyle \varGamma=N_3/(\gamma M^2)$.

In the general case, $p$, $\rho$, $V$ and $\varTheta$ derivatives with respect to natural directions are expressed through the basic flow unevennesses $N_1$, $N_2$, $N_3$, $N_7$ using (\ref{1.4}), (\ref{1.3}), (\ref{1.L}), (\ref{1.III}) and (\ref{1.LII}) by the following formulae:\pagebreak[0]
\begin{equation}
\left \{ \begin{array}{l}
\displaystyle \frac{\upartial p}{\upartial \ell}=pN_1;\\[0.5cm]
\displaystyle \frac{\upartial \rho}{\upartial \ell}=-\frac{s_p}{s_\rho}pN_1;\\[0.5cm]
\displaystyle \frac{\upartial V}{\upartial \ell}=-\frac{p}{\rho V}N_1;\\[0.5cm]
\displaystyle \frac{\upartial \varTheta}{\upartial \ell}=N_2;
\end{array}
\right.
\begin{array}{l}
\displaystyle \frac{\upartial p}{\upartial n}=-\rho V^2N_2;\\[0.5cm]
\displaystyle \frac{\upartial \rho}{\upartial n}=-\frac{p_0}{\rho_0}\cdot \frac{s_{\rho 0}}{h_{\rho 0}s_\rho}(N_3-N_7)+\frac{s_p\rho V^2}{s_\rho}N_2;\\[0.5cm]
\displaystyle \frac{\upartial V}{\upartial n}=\frac{p_0}{\rho_0V}N_7+\frac{p_0}{\rho_0V}\cdot \frac{h_\rho s_{\rho 0}}{h_{\rho 0}s_\rho}(N_3-N_7)+VN_2;\\[0.5cm]
\displaystyle \frac{\upartial \varTheta}{\upartial n}=\frac{p}{\rho V^2}N_1 +\frac{1}{\rho}\cdot\frac{s_p}{s_\rho}pN_1- \frac{\delta}{y}\sin \varTheta.
\end{array}\label{1.7}
\end{equation}

Let us also give the expressions for the basic gas flow unevennesses through the derivatives of $p$, $\rho$, $V$ and $\varTheta$ with respect to natural directions:\pagebreak[0]
\begin{equation}
\begin{array}{l}
\displaystyle N_1=\frac{\upartial \ln p}{\upartial\ell}, \quad
N_2=\frac{\upartial \varTheta}{\upartial\ell},\quad
N_7=\frac{\rho_0}{p_0}\left(h_p\frac{\upartial p}{\upartial n}+h_\rho\frac{\upartial \rho}{\upartial n}+ V \frac{\upartial V}{\upartial n}\right),\\[0.5cm]
\displaystyle N_3=\frac{\rho_0}{p_0} \left(V \frac{\upartial V}{\upartial n}+\left (h_\rho -h_{\rho0} \frac{s_\rho}{s_{\rho0}} \right) \frac{\upartial \rho}{\upartial n}+ \left(h_p -h_{\rho0} \frac{s_p}{s_{\rho0}} \right) \frac{\upartial p}{\upartial n} \right).
\end{array} \label{1.31}
\end{equation}

\section{Solution of the first--order problem for a shock wave} \label{3}\nopagebreak

In this section the basic unevennesses of the gas flow downstream a shock wave are determined. The gasdynamic parameters on both sides of the shock are known from the zero--order problem solution. The unevennesses of the flow upstream the shock and the shock surface curvatures should be prescribed. The examples of these unevennesses calculation in the thermodynamically perfect gas model and the thermally perfect, calorically imperfect gas model are presented. In the latter model the gas heat capacity depends on its temperature.

The DDCC on a shock wave in the form (\ref{1.XX}) allow to interrelate the basic unevennesses of the gas flows on the sides of the shock. This system is written in terms of $p$, $\rho$, $V$, $\varTheta$ derivatives with respect to the tangential to the shock direction $\vec \tau$. These derivatives depend on the distribution of gasdynamic parameters as well as on the geometry of the shock surface. Let us express them in terms of the derivatives with respect to natural directions and the shock parameters.

For any (scalar or vector) gasdynamic parameter $f$, defined in the flow field in the vicinity of the shock surface ($p$, $\rho$, $\vec V$, etc.), the following relations between the derivatives with respect to directions hold (Lin \& Rubinov, 1948; Eckert, 1975): \pagebreak[0]
\begin{equation}
\frac{\upartial f}{\upartial \tau}=\cos \sigma \frac{\upartial f}{\upartial \ell} + \chi \sin \sigma \frac{\upartial f}{\upartial n}; \label{1.8}
\end{equation}
\begin{equation}
\frac{\upartial \widehat f}{\upartial \tau}=\cos (\sigma-\beta) \frac{\upartial \widehat f}{\upartial \ell} + \chi \sin (\sigma-\beta) \frac{\upartial \widehat f} {\upartial n}. \label{1.12}
\end{equation}
The relation (\ref{1.8}) is valid in the upstream flow, the relation (\ref{1.12})~--- in the downstream flow. The tilde sign in the derivatives with respect to $\widehat {\vec \ell}$, $\widehat {\vec n}$ writing is omitted. Let us note that the derivatives with respect to natural directions on the shock surface are understood in the sense of one--sided. The derivatives of gasdynamic parameters with respect to the tangential directions $\vec \tau$ and $\vec b$ in the interior points of the discontinuity surface are continuous. The boundaries of the discontinuity surface correspond to the lines of its interference with other gasdynamic discontinuities or lie on solid surfaces. At these boundaries the derivatives with respect to $\vec \tau$ and $\vec b$ are understood as one--sided and they might turn to infinity.

Derivatives with respect to the tangent direction $\vec \tau$ in (\ref{1.XX}) can be expressed through the derivatives with respect to the natural directions $\vec \ell$, $\vec n$ using (\ref{1.8})--(\ref{1.12}), and they, in turn,~--- through the basic flow unevennesses using (\ref{1.7}). Since the basic unevennesses of the flow upstream the shock and the shock curvatures are prescibed, the left sides of (\ref{1.XX}) are known. Let us solve the algebraic system (\ref{1.XX}) with respect to $\upartial \widehat{\vphantom{V}p} / \upartial \tau$, $\upartial \widehat{\vphantom{V}\rho} / \upartial \tau$, $\upartial \widehat V / \upartial \tau$ and $\upartial \widehat \varTheta / \upartial \tau$. Its determinant is\pagebreak[0]
\begin{equation}
\det \left(\begin{array}{cccc}
0&\widehat{u}_\nu& \widehat{\rho} \sin (\sigma-\beta)&-\chi \widehat{\rho} u_\tau\\
0&0&\cos (\sigma-\beta)&\chi \widehat {u}_\nu\\
1&\widehat{u}_\nu^2&2 \widehat{\rho} \widehat{u}_\nu \sin (\sigma-\beta)&-2 \chi \widehat{\rho} \widehat{u}_\nu u_\tau\\
\widehat{h}_p&\widehat{h}_\rho&\widehat V&0
\end{array}\right)=\label{1.CX}
\end{equation}
$$
=\det \left(\begin{array}{cccc}
0&\widehat{u}_\nu& \widehat{\rho} \sin (\sigma-\beta)&-\chi \widehat{\rho} u_\tau\\
0&0&\cos (\sigma-\beta)&\chi \widehat {u}_\nu\\
1&-\widehat{u}_\nu^2&0&0\\
\widehat{h}_p&\widehat{h}_\rho&\widehat V&0
\end{array}\right)=
$$
$$
=\det \left(\begin{array}{cc}
1&-\widehat{u}_\nu^2\\
\widehat{h}_p&\displaystyle \widehat{h}_\rho-\frac{\widehat{u}_\nu^2}{\widehat{\rho}}
\end{array}\right)\times \chi \widehat{\rho}\det \left(\begin{array}{cc}
\sin (\sigma-\beta)&-u_\tau\\
\cos (\sigma-\beta)&\widehat {u}_\nu
\end{array}\right)=\chi \Delta\widehat {V},
$$
here (\ref{1.IL}) and the notation $\displaystyle \Delta=\widehat{\rho}\widehat{h}_\rho-(1-\widehat{\rho}\widehat{h}_p)\widehat{u}_\nu^2$ was used. Using (\ref{C.8}), the last expression can be transformed to\pagebreak[0]
\begin{equation}
\Delta=\widehat{\rho}\widehat{h}_\rho\left(1-\frac{\widehat{u}_\nu^2}{\widehat a^2}\right), \qquad \widehat{h}_\rho<0.\label{1.CVI}
\end{equation}
In (Chernyi 1994) the inequality $\widehat{u}_\nu\leq \widehat a$ is proved, and the equality \mbox{$\widehat{u}_\nu=\widehat a$} takes place if and only if the shock degenerates into a discontinuous characteristic. For a non--degenerate shock $\Delta<0$, the determinant (\ref{1.CVI}) is non--zero, and (\ref{1.XX}) has a unique solution.

At each point of the shock degeneration into a discontinuous characteristic the equalities\pagebreak[0]
\begin{equation}
\Delta=0, \quad u_\nu=\widehat{u}_\nu=a=\widehat a, \quad p=\widehat p, \quad \rho=\widehat \rho, \quad V=\widehat V, \quad \beta=0, \quad \varTheta=\widehat \varTheta\label{1.CVII}
\end{equation}
are valid. As a result, the dependence on $S_a$ disappears from (\ref{1.XX}). The equations (\ref{1.XX}) in this case form a homogeneous system of equations with respect to the differences of derivatives $\displaystyle \frac{\upartial \widehat{p}}{\upartial \tau}-\frac{\upartial p}{\upartial \tau}$, $\displaystyle \frac{\upartial \widehat{\rho}}{\upartial \tau}-\frac{\upartial \rho}{\upartial \tau}$, $\displaystyle \frac{\upartial \widehat{V}}{\upartial \tau}-\frac{\upartial V}{\upartial \tau}$ and $\displaystyle \frac{\upartial \widehat{\varTheta}}{\upartial \tau}-\frac{\upartial \varTheta}{\upartial \tau}$, and the determinant of this system is equal to (\ref{1.CX}) and is zero in case of a discontinuous characteristic. Thus, at each point of the shock degeneration (\ref{1.XX}) has an infinite number of solutions.

On a discontinuous characteristic surface the relations (\ref{1.CVII}) can be differentiated with respect to the tangential direction $\vec \tau$, so that
\begin{equation}
\frac{\upartial \widehat{p}}{\upartial \tau}-\frac{\upartial p}{\upartial \tau}=0, \qquad \frac{\upartial \widehat{\rho}}{\upartial \tau}-\frac{\upartial \rho}{\upartial \tau}=0, \qquad \frac{\upartial \widehat{V}}{\upartial \tau}-\frac{\upartial V}{\upartial \tau}=0, \qquad \frac{\upartial \widehat{\varTheta}}{\upartial \tau}-\frac{\upartial \varTheta}{\upartial \tau}=0.
\label{1ast}
\end{equation}
The equalities (\ref{1ast}) give a trivial solution of the system of equations (\ref{1.XX}); this solution takes place on the discontinuous characteristic surface.

In case the shock degenerates into a discontinuous characteristic only on a straight line (circle) that is perpendicular to the flow plane (meridional half--plane), the equalities (\ref{1.CVII}) cannot be differentiated with respect to $\vec \tau$. Such situation takes place, for example, on the line of origin of a hanging shock wave in jet flows, or in the centre of a centered compression wave. The determination of the derivatives differences values on the sides of such discontinuities is yet an unsolved problem.

For non--degenerate shocks the solution of (\ref{1.XX}) has the form:\pagebreak[0]
\begin{equation}
\hspace{-1.49cm}\begin{array}{l}
\displaystyle \frac{\upartial \widehat{p}}{\upartial \tau}=\left [ \left (\widehat \rho \widehat{h}_\rho-\widehat u_\nu^2+\widehat \rho \widehat u_\nu^2 h_p\right) \frac{\upartial p}{\upartial \tau}+\right.\\[0.5cm]
\displaystyle \qquad +\left (\left(\widehat \rho \widehat{h}_\rho-\widehat u_\nu^2\right)u_\nu(u_\nu-2\widehat{u}_\nu)+\widehat u_\nu^2(\widehat \rho h_\rho-\widehat u_\nu u_\nu) \right) \frac{\upartial \rho}{\upartial \tau}+\\[0.5cm]
\displaystyle \quad+(u_\nu-\widehat{u}_\nu) \left (\left(\widehat \rho \widehat{h}_\rho-\widehat u_\nu^2\right)2\rho +\widehat \rho \widehat u_\nu^2\frac{u_\nu+\widehat{u}_\nu}{u_\nu}\right)\left(\sin \sigma \frac{\upartial V}{\upartial \tau}+u_\tau S_a -\chi u_\tau \frac{\upartial \varTheta}{\upartial \tau}\right)\biggr] \Delta^{-1},
\end{array}\hspace{-1.49cm}\label{1.CI}
\end{equation}
\begin{equation}
\begin{array}{l}
\displaystyle \frac{\upartial \widehat \rho}{\upartial \tau}=\left [\widehat \rho (h_p- \widehat{h}_p)\frac{\upartial p}{\upartial \tau}+ \left(-\widehat{\rho} \widehat{h}_p u_\nu(u_\nu-2\widehat{u}_\nu)+ \widehat{\rho} h_\rho-\widehat{u}_\nu u_\nu \right) \frac{\upartial \rho}{\upartial \tau} + \right.\\[0.5cm]
\displaystyle + \widehat \rho (u_\nu-\widehat{u}_\nu) \left (\frac{u_\nu+\widehat{u}_\nu}{u_\nu}- 2\widehat{h}_p\rho\right)\left(\sin \sigma \frac{\upartial V}{\upartial \tau}+u_\tau S_a
-\chi u_\tau \frac{\upartial \varTheta}{\upartial \tau}\right)\biggr] \Delta^{-1},
\end{array}\hspace{-0.3cm}\label{1.CII}
\end{equation}
\begin{equation}
\begin{array}{l}
\displaystyle \frac{\upartial \widehat V}{\upartial \tau}=\biggl[(\widehat{h}_p-h_p)\widehat{u}_\nu \sin (\sigma-\beta) \frac{\upartial p}{\upartial \tau}+\\[0.5cm]
\displaystyle \qquad +\left(u_\nu\left(\widehat{h}_\rho+\widehat{h}_p\widehat{u}_\nu(u_\nu-\widehat{u}_\nu)\right)-\widehat{u}_\nu h_\rho \right)\sin (\sigma-\beta) \frac{\upartial \rho}{\upartial \tau}+\\[0.5cm]
\displaystyle \qquad+\frac{\widehat V}{V}\Delta\frac{\upartial V}{\upartial \tau}-\frac{u_\tau(u_\nu-\widehat{u}_\nu)(u_\nu+\widehat{u}_\nu)}{\widehat V u_\nu}\Delta\left(S_a-\chi \frac{\upartial \varTheta}{\upartial \tau}\right)+\\[0.5cm]
\displaystyle \quad +\widehat{u}_\nu (u_\nu-\widehat{u}_\nu) \left(2\widehat{h}_p \rho-\frac{u_\nu+\widehat{u}_\nu}{u_\nu}\right)\sin (\sigma-\beta)\left(\sin \sigma \frac{\upartial V}{\upartial \tau}+u_\tau S_a-\chi u_\tau \frac{\upartial \varTheta}{\upartial \tau}\right)\biggr] \Delta^{-1},
\end{array}\hspace{-1.0cm}\label{1.CIII}
\end{equation}
\begin{equation}
\begin{array}{l}
\displaystyle \frac{\upartial \widehat \varTheta}{\upartial \tau}=\chi S_a-\chi \biggl[(\widehat{h}_p-h_p)\sin (\sigma-\beta)\cos(\sigma-\beta) \frac{\upartial p}{\upartial \tau}+\\[0.5cm]
\displaystyle \qquad +\frac{1}{\widehat V}\left(u_\nu\left(\widehat{h}_\rho+\widehat{h}_p\widehat{u}_\nu(u_\nu-\widehat{u}_\nu)\right)-\widehat{u}_\nu h_\rho \right)\cos(\sigma-\beta)\frac{\upartial \rho}{\upartial \tau}+\\[0.5cm]
\displaystyle \qquad +\frac{V^2 \widehat{u}_\nu}{\widehat V^2 u_\nu}\Delta\left(S_a-\chi \frac{\upartial \varTheta}{\upartial \tau}\right)+(u_\nu-\widehat{u}_\nu) \left (2\widehat{h}_p \rho-\frac{u_\nu+\widehat{u}_\nu}{u_\nu}\right)\times\\[0.5cm]
\displaystyle \qquad \qquad \times\sin (\sigma-\beta) \cos (\sigma-\beta) \left(\sin \sigma \frac{\upartial V}{\upartial \tau}+u_\tau S_a -\chi u_\tau \frac{\upartial \varTheta}{\upartial \tau}\right)\biggr] \Delta^{-1}.
\end{array}\hspace{-1.3cm}\label{1.CIV}
\end{equation}

Let us express the derivative $\partial \widehat p/\partial\ell$ behind a shock wave through the derivatives with respect to $\vec \tau$. Let us multiply (\ref{1.30a}) by $\chi \sin (\sigma-\beta)$, (\ref{1.30c}) by $\cos (\sigma-\beta)$, add them together, substitute $\partial \widehat d/\partial\ell$ from (\ref{1.30b}) and express the parameters derivatives with respect to the normal $\vec n$ through their derivatives with respect to $\vec \tau$ using (\ref{1.12}), we get:\pagebreak[0]
$$
2\chi\left(\frac{\widehat V^2}{2\widehat a^2}-1\right) \frac{\upartial \widehat p}{\upartial\ell} \sin (\sigma-\beta)+2\widehat d\frac{\upartial \widehat \varTheta}{\upartial \tau}+2\chi N_4 \widehat d\sin \widehat \varTheta\sin (\sigma-\beta)=
$$
$$
=-\chi \sin (\sigma-\beta)\frac{\upartial \widehat p}{\upartial\ell}-\cos(\sigma-\beta)\frac{\upartial \widehat p}{\upartial n}=-\chi \cot(\sigma-\beta)\frac{\upartial \widehat p}{\upartial\tau}+\chi \left(\frac{\cos^2(\sigma-\beta)}{\sin (\sigma-\beta)}-\sin (\sigma-\beta)\right)\frac{\upartial \widehat p}{\upartial\ell},
$$
As a result, using (\ref{1.IL}) and (\ref{hTheta}), we have\pagebreak[0]
\begin{equation}
\begin{array}{l}
\displaystyle \left(1-\frac{\widehat{u}^2_\nu}{\widehat a^2}\right)\frac{\upartial \widehat p}{\upartial\ell}=\chi \widehat \rho \widehat V^2\frac{\upartial \widehat \varTheta}{\upartial \tau}\sin(\sigma-\beta)+N_4 \widehat \rho \widehat{u}^2_\nu\sin (\varTheta+\chi\beta)+\cos(\sigma-\beta)\frac{\upartial \widehat p}{\upartial\tau}.\end{array}\label{1.CV}
\end{equation}

Let us describe an algorithm for the gas flow basic unevennesses $\widehat{N}_1$, $\widehat{N}_2$, $\widehat{N}_3$, $\widehat{N}_7$ calculation for a non--degenerate shock wave through the gasdynamic parameters, the gas flow before the shock basic unevennesses $N_1$, $N_2$, $N_3$, $N_7$, the shock direction index $\chi$ and its curvatures $N_4$ and $S_a$:

--- the derivatives of the main gasdynamic parameters upstream the shock with respect to the natural directions are determined by formulae (\ref{1.7});

--- the transition to the derivatives with respect to $\vec \tau$ is made using (\ref{1.8});

--- the derivatives of the main gasdynamic parameters downstream the shock wave with respect to $\vec \tau$ are calculated using the solution (\ref{1.CI})--(\ref{1.CIV}) of the system (\ref{1.XX});

--- the transition back to the derivatives with respect to the natural directions is held: the pressure derivative $\upartial \widehat p / \upartial \ell$ is calculated from (\ref{1.CV}), the derivatives $\upartial \widehat V / \upartial \ell$, $\upartial \widehat {\vphantom{V}\rho} / \upartial \ell$, $\upartial \widehat \varTheta / \upartial n$ are determined by (\ref{1.3b}), (\ref{1.3d}) and (\ref{1.3a}), respectively, $\upartial \widehat {\vphantom{V}p} / \upartial n$, $\upartial \widehat {\vphantom{V}\rho} / \upartial n$, $\upartial \widehat V / \upartial n$~--- by formulae (\ref{1.12}), and $\upartial \widehat \varTheta / \upartial \ell$~--- from (\ref{1.3c});

--- the basic unevennesses of the gas flow behind the shock are calculated by the formulae (\ref{1.31}).

All the used relations are linear and homogeneous with respect to the gasdynamic parameters derivatives and the shock wave curvatures. The interrelations between derivatives with respect to different directions are linear and homogeneous as well. Consequently, the gas flow downstream the shock basic unevennesses are linear homogeneous functions of the gas flow upstream the shock basic unevennesses $N_1$, $N_2$, $N_3$, $N_7$ and the shock curvatures $S_a$ and $N_4$. For them, the expansions
\begin{equation}
\widehat N_i=\mathsfbi{A}_{i1}N_1+\mathsfbi{A}_{i2}N_2+\mathsfbi{A}_{i3}N_3 +\mathsfbi{A}_{i4}N_4+\mathsfbi{A}_{i5}S_a+\mathsfbi{A}_{i7}N_7, \quad i=1, \, 2, \, 3, \, 7, \label{Aij}
\end{equation}
are valid; here the coefficients $\mathsfbi{A}_{ij}$ depend only on the gasdynamic parameters, they are finite for any non--degenerate shock ($J>1$). Each of the influence coefficients $\mathsfbi{A}_{ij}$ characterizes the dependence of the unevenness $\widehat N_i$ on one of the input parameters $N_1$, $N_2$, $N_3$, $N_7$, $N_4$ and $S_a$, under the condition that the remaining five parameters are zero. In plane flows the curvature $N_4=0$, and the coefficients $\mathsfbi{A}_{i4}$ are not defined. All other coefficients $\mathsfbi{A}_{ij}$ for plane and axisymmetric flows are the same. The expansion of the form (\ref{Aij}) was used before in (Uskov 1983) and (Adrianov \etal 1995).

Let us note that the non--isoenthalpy factor $\widehat N_7$ of the flow downstream the shock wave depends only on the non--isoenthalpy factor $N_7$. Really, since the rest enthalpy $h_0$ is constant along the streamline and is the same on the sides of the shock, using the relations (\ref{1.8}) and (\ref{1.12}) between derivatives we can obtain the equalities
$$
\widehat{N}_7=\frac{\widehat \rho_0}{\widehat p_0}\frac{\upartial \widehat h_0}{\upartial n}=\chi \frac{\widehat \rho_0}{\widehat p_0 \sin(\sigma-\beta)}\frac{\upartial \widehat h_0}{\upartial \tau}=
\chi \frac{\widehat \rho_0}{\widehat p_0 \sin(\sigma-\beta)}\frac{\upartial h_0}{\upartial \tau}=\frac{\widehat \rho_0\sin\sigma}{\widehat p_0 \sin(\sigma-\beta)}\frac{\upartial h_0}{\upartial n}.
$$
Consequently,
\begin{equation}
\mathsfbi{A}_{77}=\frac{p_0\widehat \rho_0\sin\sigma}{\widehat p_0\rho_0 \sin(\sigma-\beta)}, \qquad \mathsfbi{A}_{7j}=0 \quad \mbox{for} \quad j\neq 7.\label{hN7}
\end{equation}
For a thermally perfect gas the equality $\mathsfbi{A}_{77}=\sin \sigma/\sin (\sigma-\beta)$ is valid.

Let us find the dependence of the basic gas flow unevennesses and the coefficients $\mathsfbi{A}_{ij}$ on the velocity vector polar angle $\varTheta$ and the shock direction index relative to the incident flow $\chi=\pm 1$. In plane flows the axes $x$, $y$ are chosen arbitrarily in the flow plane. If you turn the axes maintaining their relative orientation, the polar angles $\varTheta$ and $\widehat \varTheta$ vary by a constant item, the rest gasdynamic parameters and its derivatives remain constant. Consequently, in plane flows the coefficients $\mathsfbi{A}_{ij}$ are independent of the angle $\varTheta$. The sign of $\chi$, as stated above, is determined by the direction of the $y$ axis relative to the $x$ axis. If it is changed to the opposite one, the signs of the polar angles $\varTheta$, $\widehat \varTheta$, $\varOmega$ change and the direction of the vector $\vec n$ changes either; the formulae (\ref{1.XXIV}), (\ref{1.4}) and (\ref{1.I}) show that in this case $N_1$ and the curvature $S_a$ remain their values, and $N_2$, $N_3$ and $N_7$ change their signs. Consequently, when the sign of $\chi$ is changed the quantities $\mathsfbi{A}_{11}$, $\chi \mathsfbi{A}_{12}$, $\chi \mathsfbi{A}_{13}$, $\mathsfbi{A}_{15}$, $\chi \mathsfbi{A}_{17}$, $\chi \mathsfbi{A}_{21}$, $\mathsfbi{A}_{22}$, $\mathsfbi{A}_{23}$, $\chi \mathsfbi{A}_{25}$, $\mathsfbi{A}_{27}$, $\chi \mathsfbi{A}_{31}$, $\mathsfbi{A}_{32}$, $\mathsfbi{A}_{33}$, $\chi \mathsfbi{A}_{35}$, $\mathsfbi{A}_{37}$ remain unchanged, and they can depend only on the pressure $p$, temperature $T$ and the Mach number $M$ of the upstream flow and the shock inclination angle $\sigma$.

If all the basic unevennesses of the incoming flow and the shock curvatures are finite, the basic unevennesses of the gas flow behind the shock will also be finite quantities for all values of gasdynamic parameters. This fact is stated in (M\"older 1979). The basic unevennesses of the gas flow before the shock and its curvatures can become infinite only at the points of the shock interference with other gasdynamic discontinuities.

Let us give an example of numerical calculations. Let us consider a shock wave in a gas flow with Mach number $M=5$. The calculations were conducted in the limits of a thermodynamically perfect diatomic gas model (the specific heats ratio $\gamma=1,4$). In this case the coefficients $\mathsfbi{A}_{ij}$ (for $i \neq 7$ and $j \neq 7$) can also be calculated according to the formulae cited in (Uskov 1983) and (Adrianov \etal 1995). The values obtained in this paper and the mentioned above precisely coincide.

\begin{figure}
\vspace{-0.2cm}
\begin{center}
\mbox {\hspace{-0.5cm} \includegraphics[width=0.55 \textwidth]{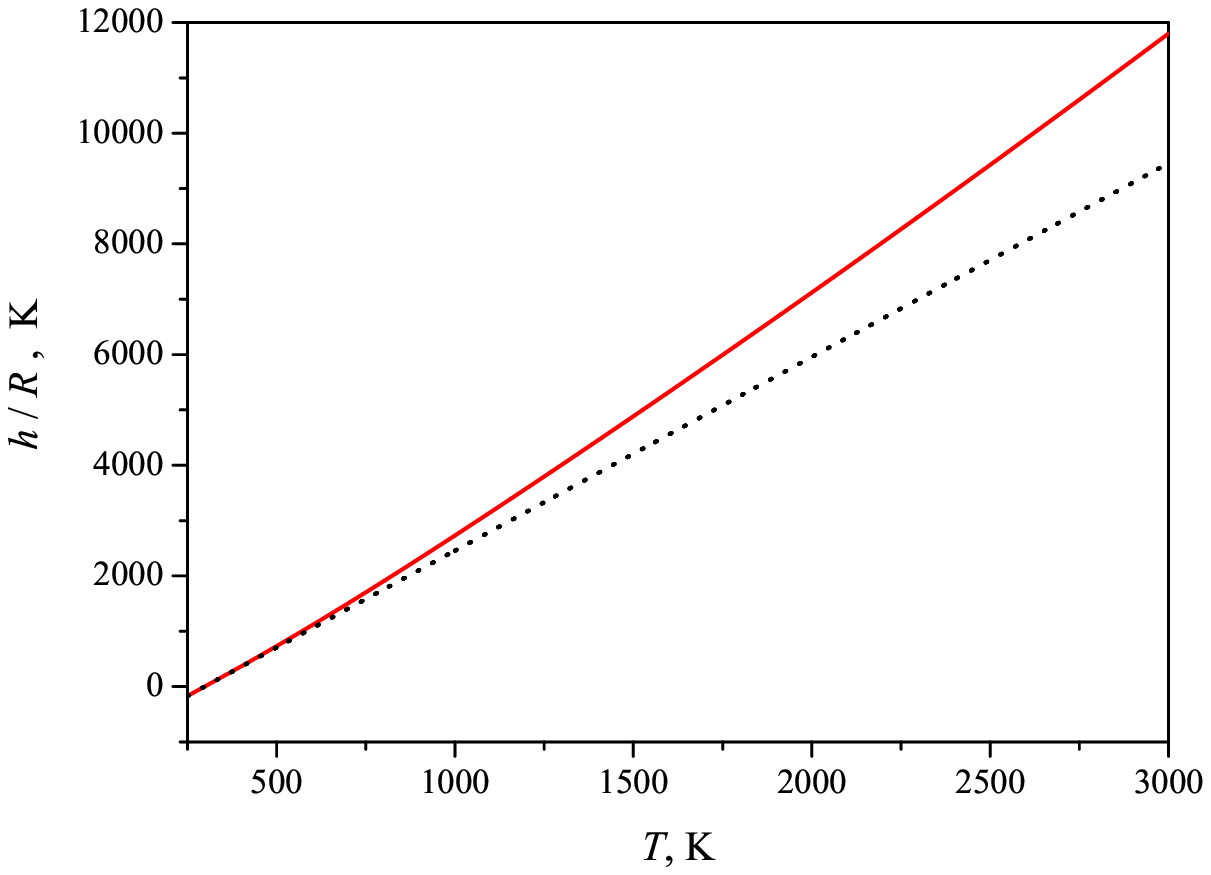} \hspace {-0.9cm} \raisebox{-1mm}{\includegraphics[width=0.55 \textwidth] {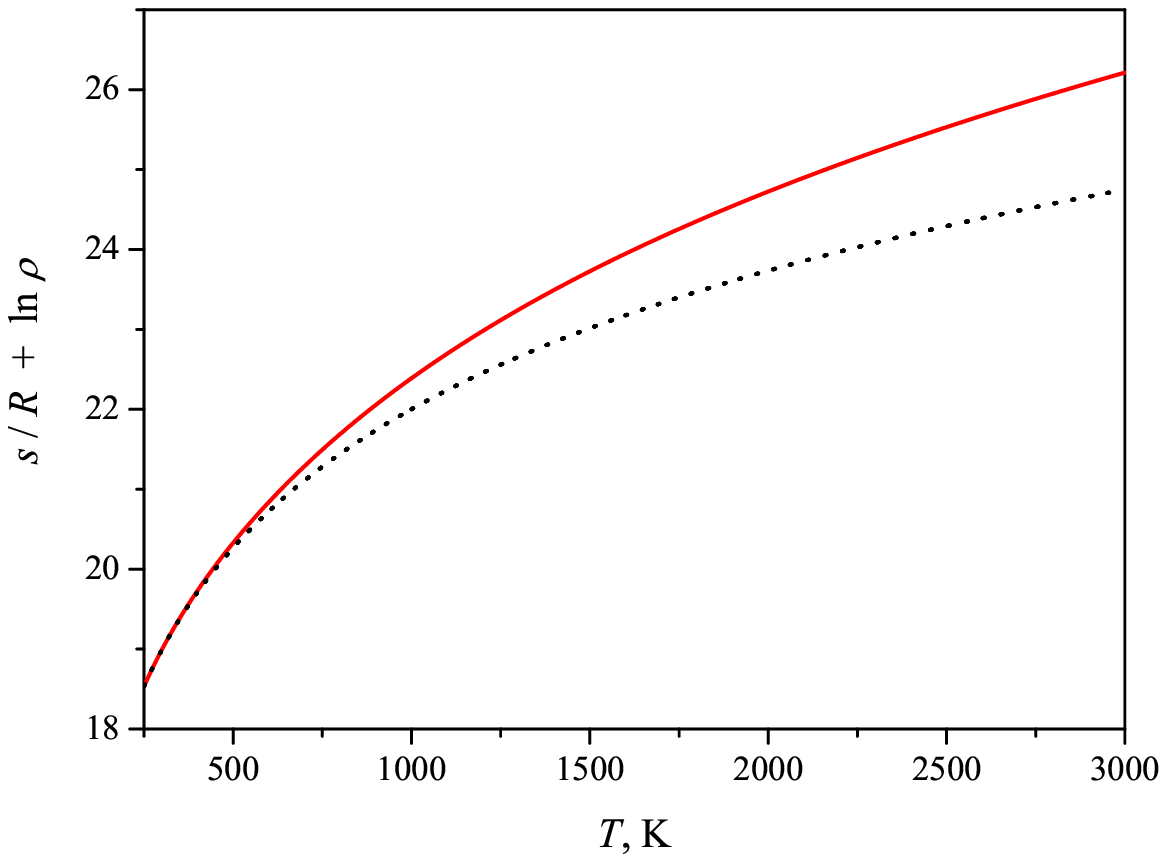}} \hspace{-0.5cm}}
\begin{picture}(0,0)
\put(-386,135){\large $(a)$}
\put(-194,135){\large $(b)$}
\end{picture}
\end{center}
\vspace{-0.5cm}
\caption{The dependences of the gas specific enthalpy and entropy on its temperature. Diatomic perfect gas~--- dotted lines; oxygen in the thermally perfect gas model~--- solid lines.}\label{SSW}
\end{figure}

The calculations were also conducted in the limits of a thermally perfect gas model. In this model the thermal Clapeyron equation of gas state is assumed valid, and the dependences of the specific enthalpy $h$ and the specific entropy $s$ are given by the following formulae:
$$
h(T)=R \left(-\frac{A_1}{T}+A_2\ln T +A_3 T +\frac{A_4}{2} T^2+\frac{A_5}{3} T^3 +\frac{A_6}{4} T^4 +\frac{A_7}{5} T^5+\frac{A_8}{6} T^6+B_1 \right);
$$
$$
s(T,\rho)=R\left(-\frac{A_1}{2T^2}-\frac{A_2}T +(A_3-1)\ln T +A_4T +\frac{A_5}{2} T^2 + \cdots+\frac{A_8}{5} T^5+B_2 -\ln \rho \right).
$$
The coefficients $R$, $A_1, \ldots, A_8$, $B_1$, $B_2$ in these expansions for different gases are cited in (McBride \etal 1963; McBride, Gordon \& Reno 1993). This model was previously considered by the authors in (Uskov \& Mostovykh 2011). In this paper, oxygen is taken as a calorically imperfect gas; a comparison of its enthalpy and entropy with a diatomic perfect gas is shown in figure~\ref{SSW}. 

\begin{figure}
\begin{center}
\mbox{\hspace{-0.5cm} \raisebox{6mm}{\includegraphics[width=0.55 \textwidth]{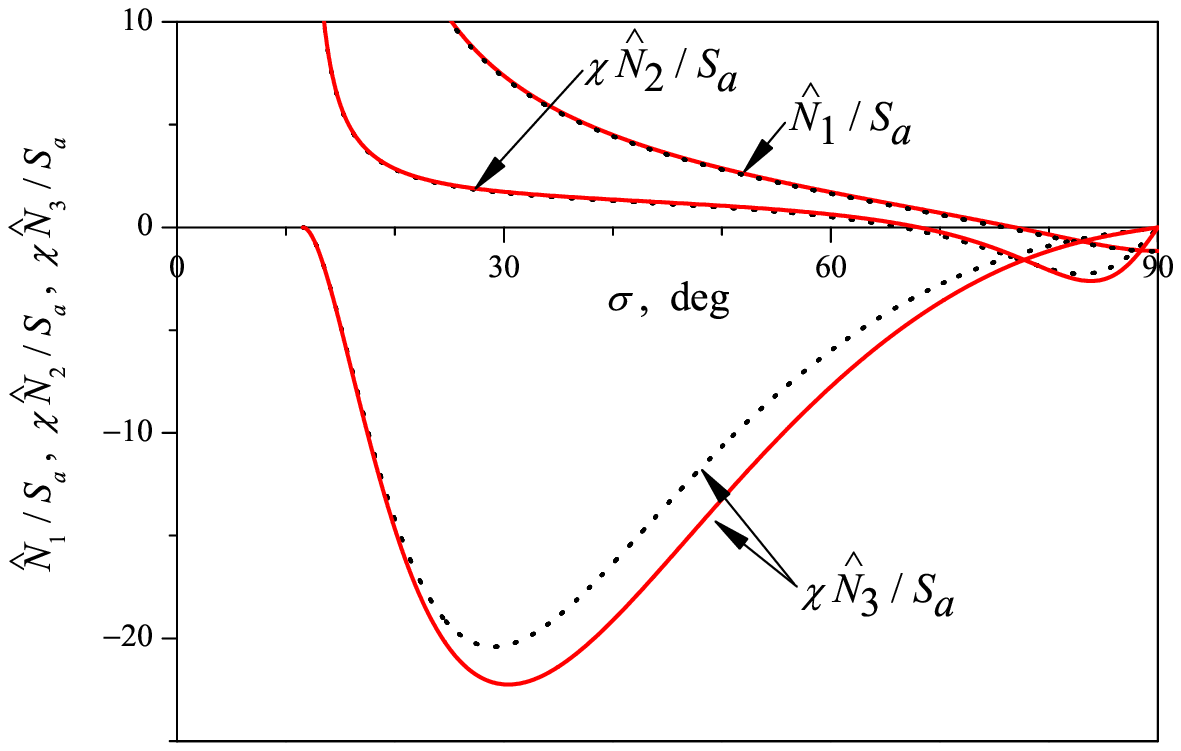}} \hspace{-0.9cm} \includegraphics[width=0.55 \textwidth]{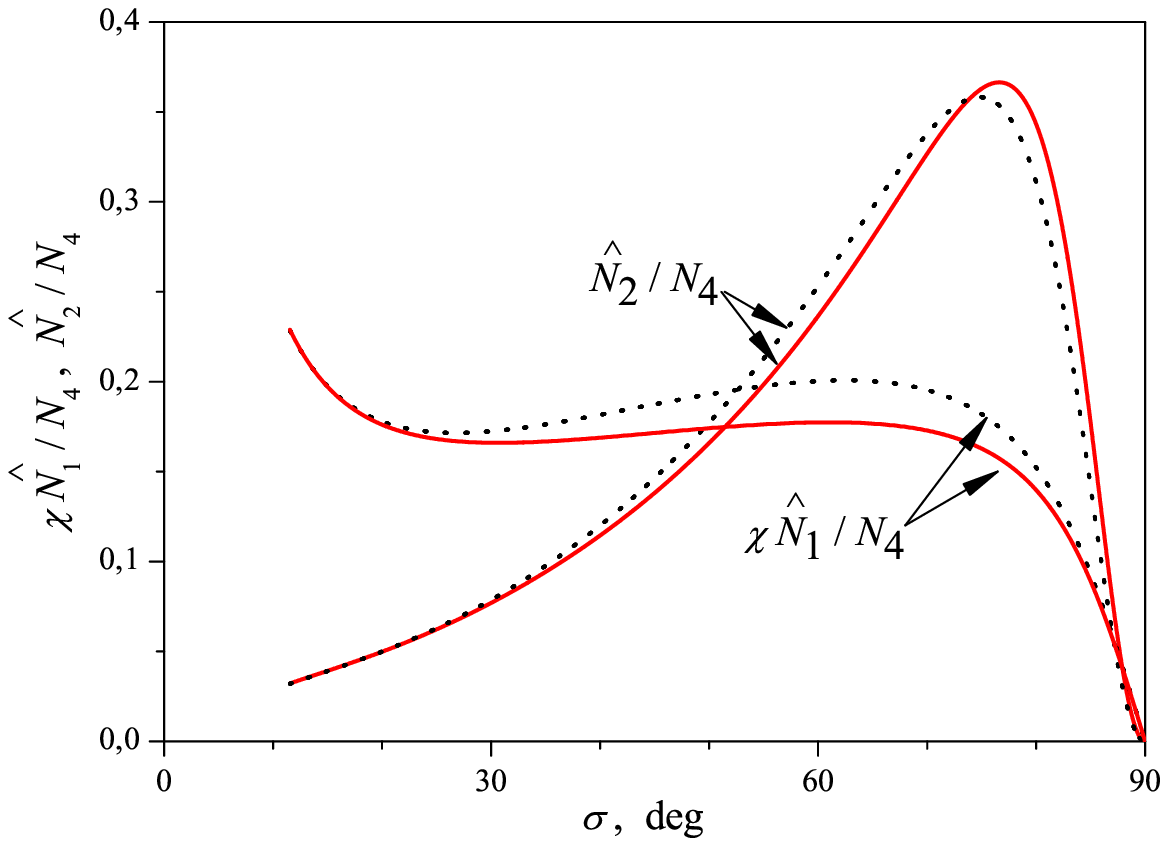} \hspace{-0.5cm}}\\
\vspace{-0.6cm}
\mbox{\hspace{-0.5cm} \includegraphics[width=0.55\textwidth]{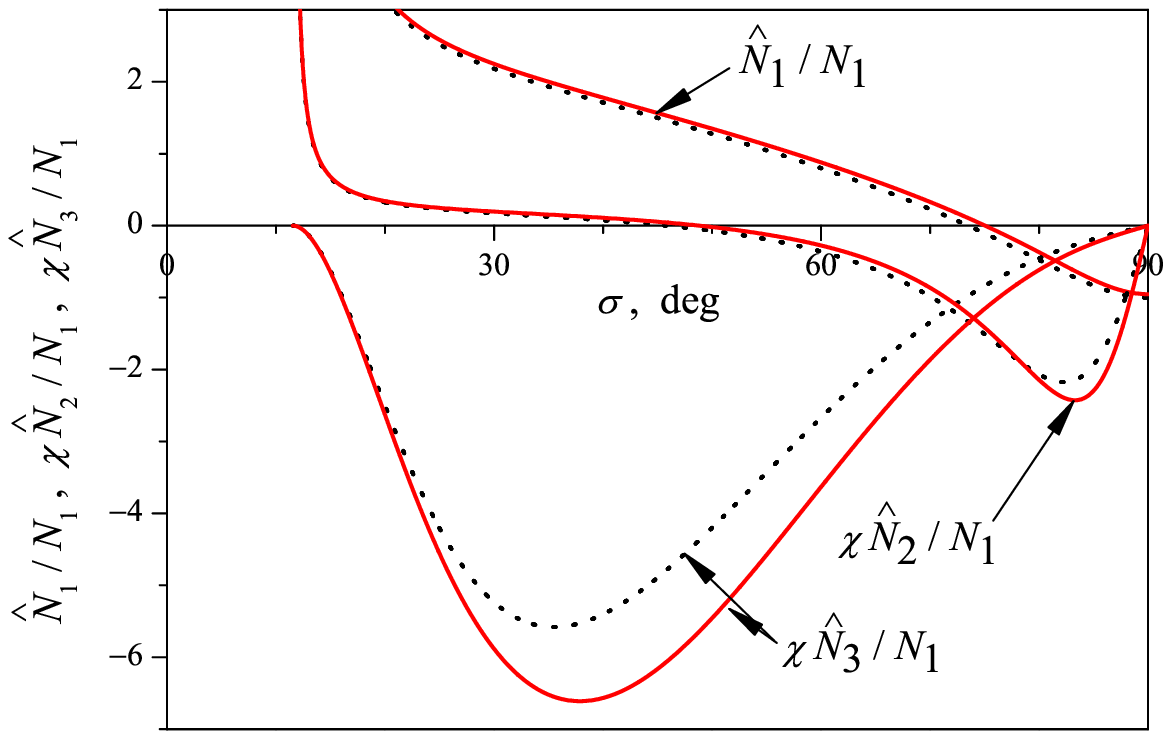} \hspace{-0.9cm} \includegraphics[width=0.55\textwidth]{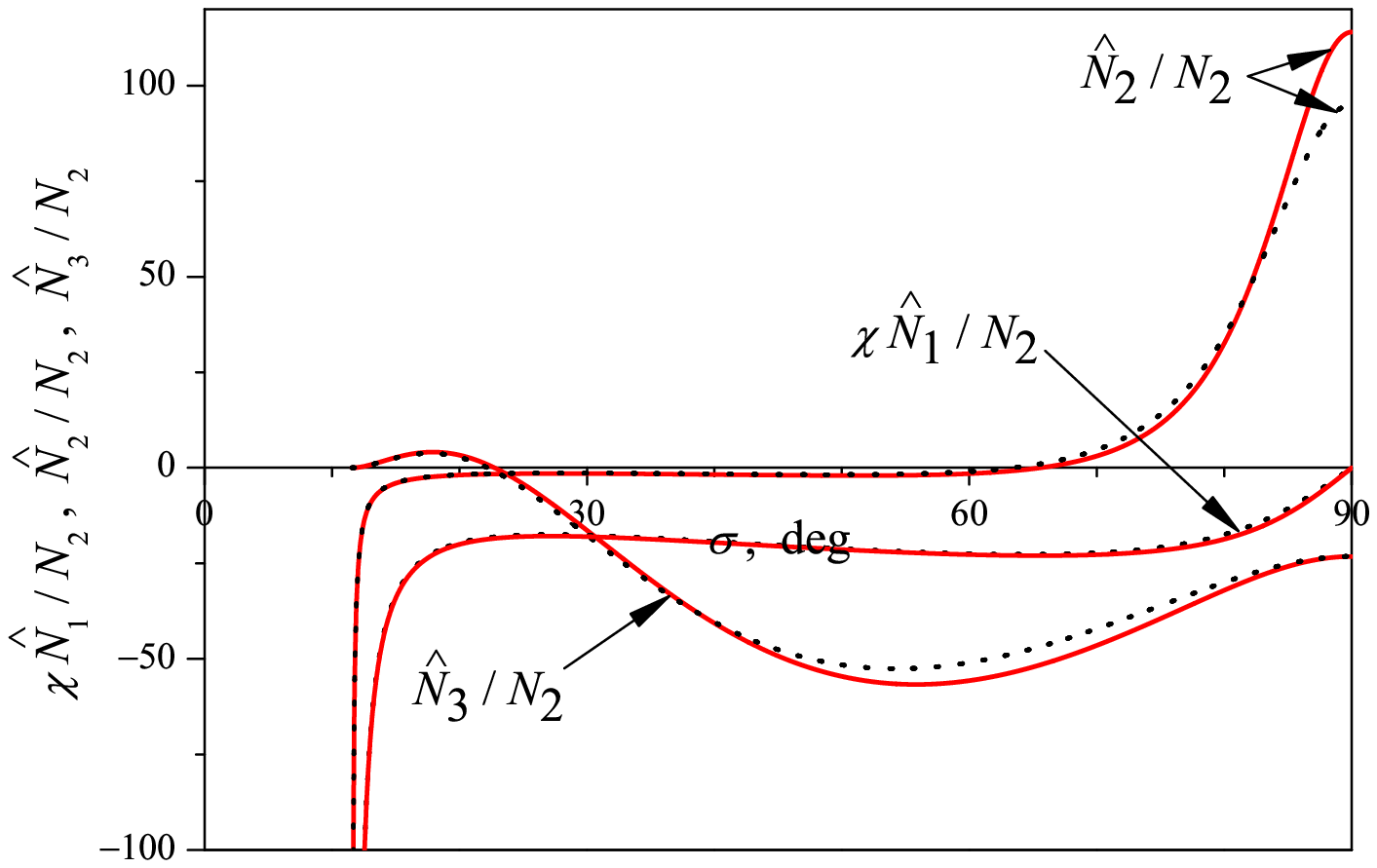} \hspace{-0.5cm}}\\
\vspace{-0.6cm}
\mbox{\hspace{-0.5cm} \includegraphics[width=0.55\textwidth]{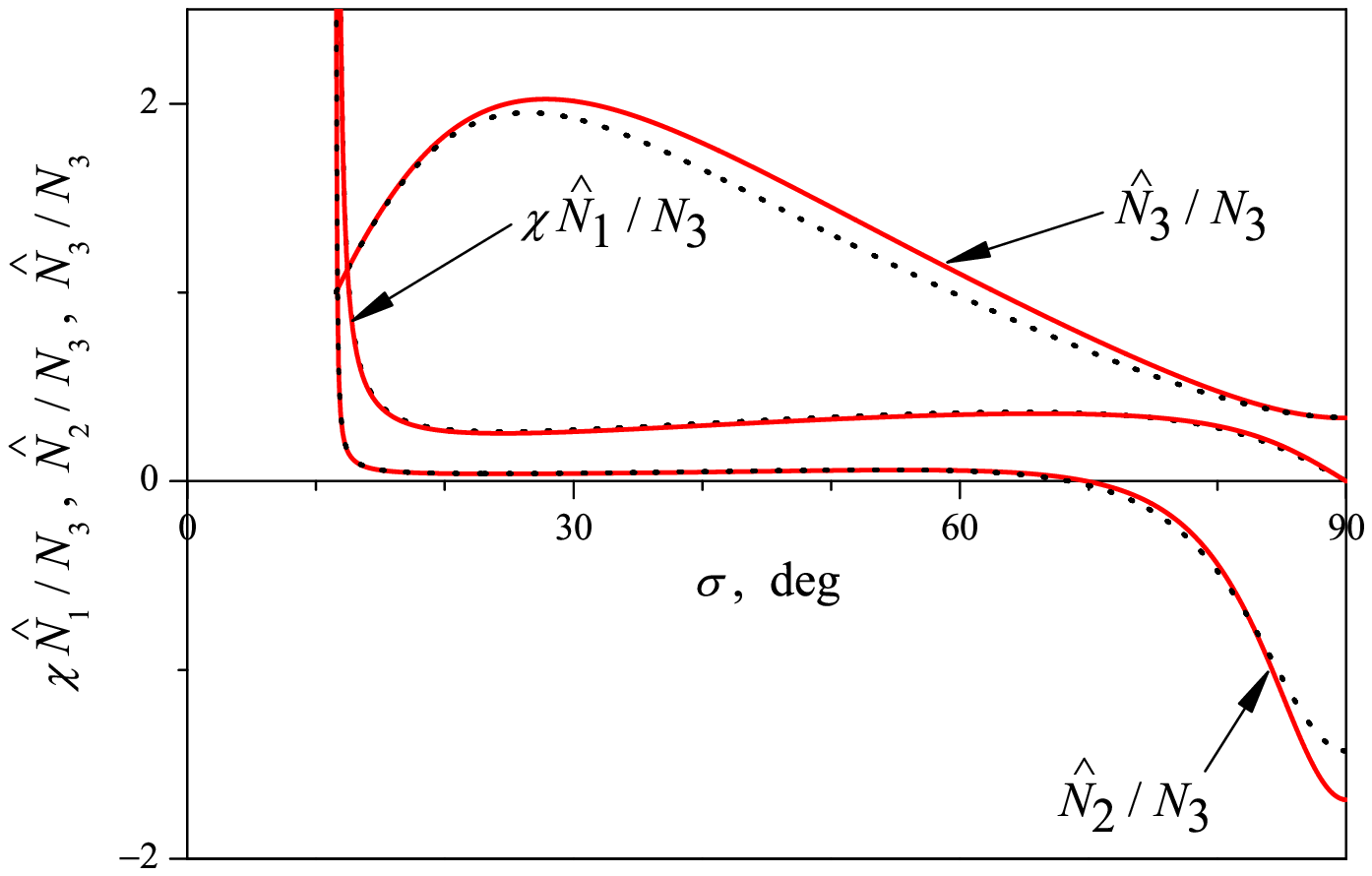} \hspace{-0.9cm} \includegraphics[width=0.55\textwidth]{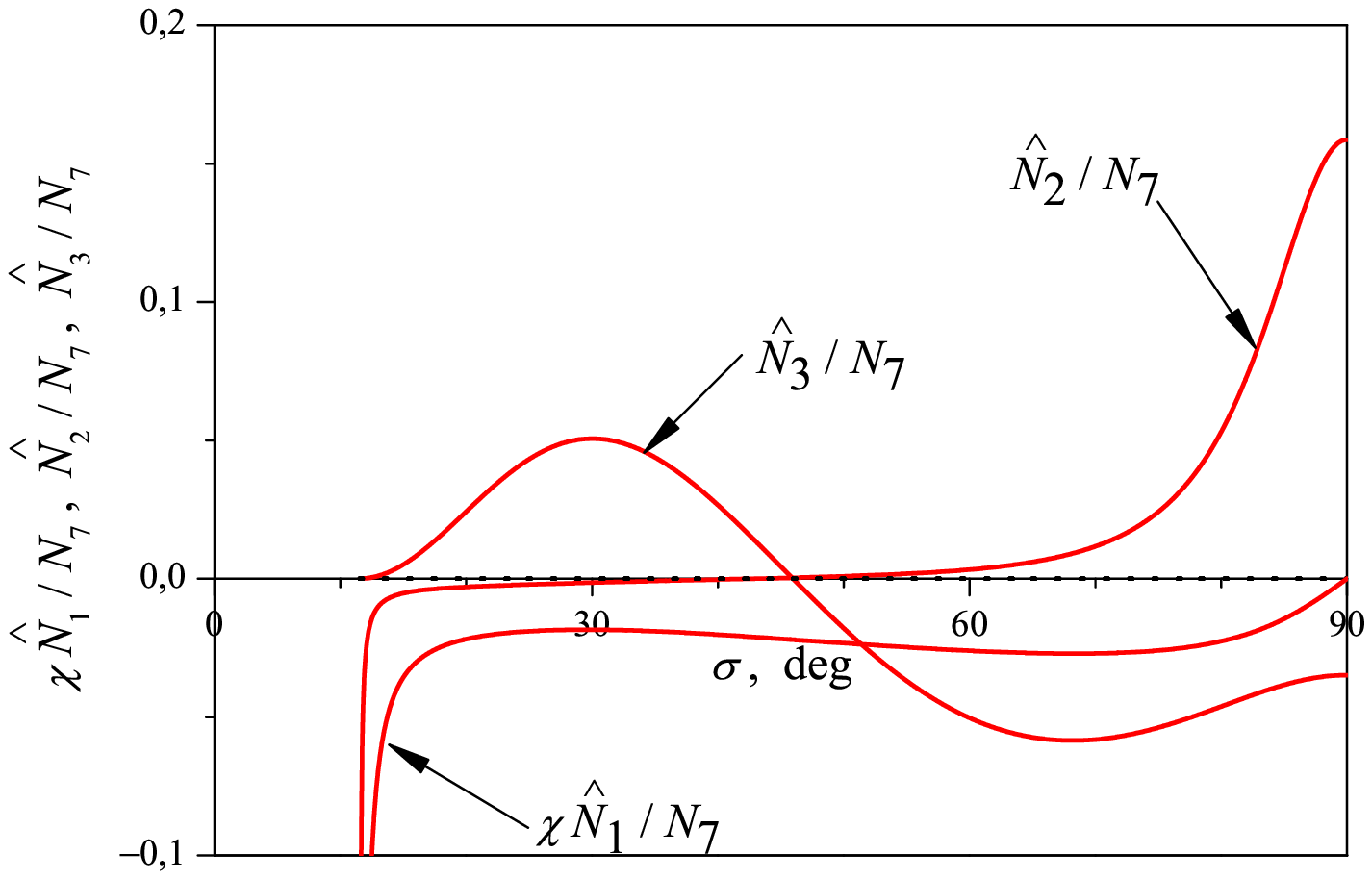} \hspace{-0.5cm}}
\begin{picture}(0,0)
\put(-380,125){\large $(e)$}
\put(-190,125){\large $(f)$}
\put(-380,250){\large $(c)$}
\put(-190,250){\large $(d)$}
\put(-380,385){\large $(a)$}
\put(-190,385){\large $(b)$}
\end{picture}

\vspace{-0.5cm}

\caption{The dependences of the influence coefficients $\mathsfbi{A}_{ij}$, $i=1, \, 2, \, 3$, $j=1, \, 2, \, 3, \, 4, \, 5, \, 7$, on the shock wave inclination angle $\sigma$. The upstream flow Mach number $M=5$, pressure $p=10^5$~Pa, temperature $T=300$~K. Diatomic perfect gas~--- black dotted lines; oxygen in the thermally perfect gas model~--- red solid lines.}\label{SSWN}

\end{center}
\end{figure}

Figure~\ref{SSWN}($a$--$b$) shows the dependences of $\mathsfbi{A}_{i5}=\widehat N_i/S_a$, $i=1, \, 2, \, 3$, for $N_4=0$ and \mbox{$\mathsfbi{A}_{i4}=\widehat N_i/N_4$} $i=1, \, 2$, for $S_a=0$ versus the shock inclination angle $\sigma$ in a uniform upstream flow. These ratios depict the effect of the shock curvatures on the basic unevennesses of the flow downstream it. The ratios $\widehat N_i/N_4$ are calculated under the assumption that the initial flow is parallel to the axis of symmetry ($\varTheta=0$). In this case $\widehat N_3/N_4=0$, i.e. the unevenness $\widehat N_3$ does not depend on the curvature $N_4$, and the quantities $\chi \widehat N_1/N_4$, $\widehat N_2/N_4$ do not depend on the index $\chi$ value.

Figure~\ref{SSWN}($c$--$f$) gives the dependences of $\mathsfbi{A}_{ij}=\widehat N_i/N_j$, $i=1, \, 2, \, 3$, $j=1, \, 2, \, 3, \, 7$ versus the shock inclination angle $\sigma$. Each of the influence coefficients $\mathsfbi{A}_{ij}$ characterizes the effect of the unevenness $N_j$ of the upstream flow on the unevenness $\widehat N_i$ of the downstream flow, under the condition that other $N_j$, $N_4$ and $S_a$ are equal to zero. The coefficients $\mathsfbi{A}_{i7}$ for a thermodynamically perfect gas are zero. This agrees with the fact that the unevennesses $N_1$, $N_2$, $N_3$ correspond to the flow describtion in the limits of a closed system of equations (\ref{1.30}). The unevenness $N_7$ is not included in that describtion.

So far, the upstream flow unevennesses and the shock curvatures $N_4$ and $S_a$ were assumed to be known. However, the proposed approach can be used for the consideration of the gasdynamic situations in which one of the basic unevennesses of the flow downstream the shock is known. The shock curvature $S_a$ in this case is determined. Examples of such flows are: the flow around a curved wall (the unevenness $\widehat N_2$ is known), and the expiration of a free jet from a nozzle. In the latter case, the flow along the boundary of the jet approximately may be considered isobaric, $\widehat N_1\approx 0$.

The gas outflow from a conical nozzle is modelled as a flow from a point source (a cylindrical source in a plane flow and a spherical source in an axisymmetric flow). In this case, the flow nonisobaric factor in the nozzle is (Uskov \& Chernyshov 2006)
$$
N_1=\frac{(1+\delta)\rho V^2}{pR\left(1-V^2/a^2\right)},
$$
here $R$ is the distance from the observation point to the center of the point source; the streamlines are straight ($N_2=0$), the gas rest parameters remain unchanged in the direction normal to the streamlines ($N_3=N_7=0$), the curvature $N_4$ is equal to
$$
N_4=\frac{\delta}{R\sin \varTheta},
$$
the velocity polar angle $\varTheta$ on the nozzle edge coincides with the nozzle opening half--angle, and the index $\chi=-1$.

\begin{figure}
\begin{center}
\includegraphics[width=0.55\textwidth]{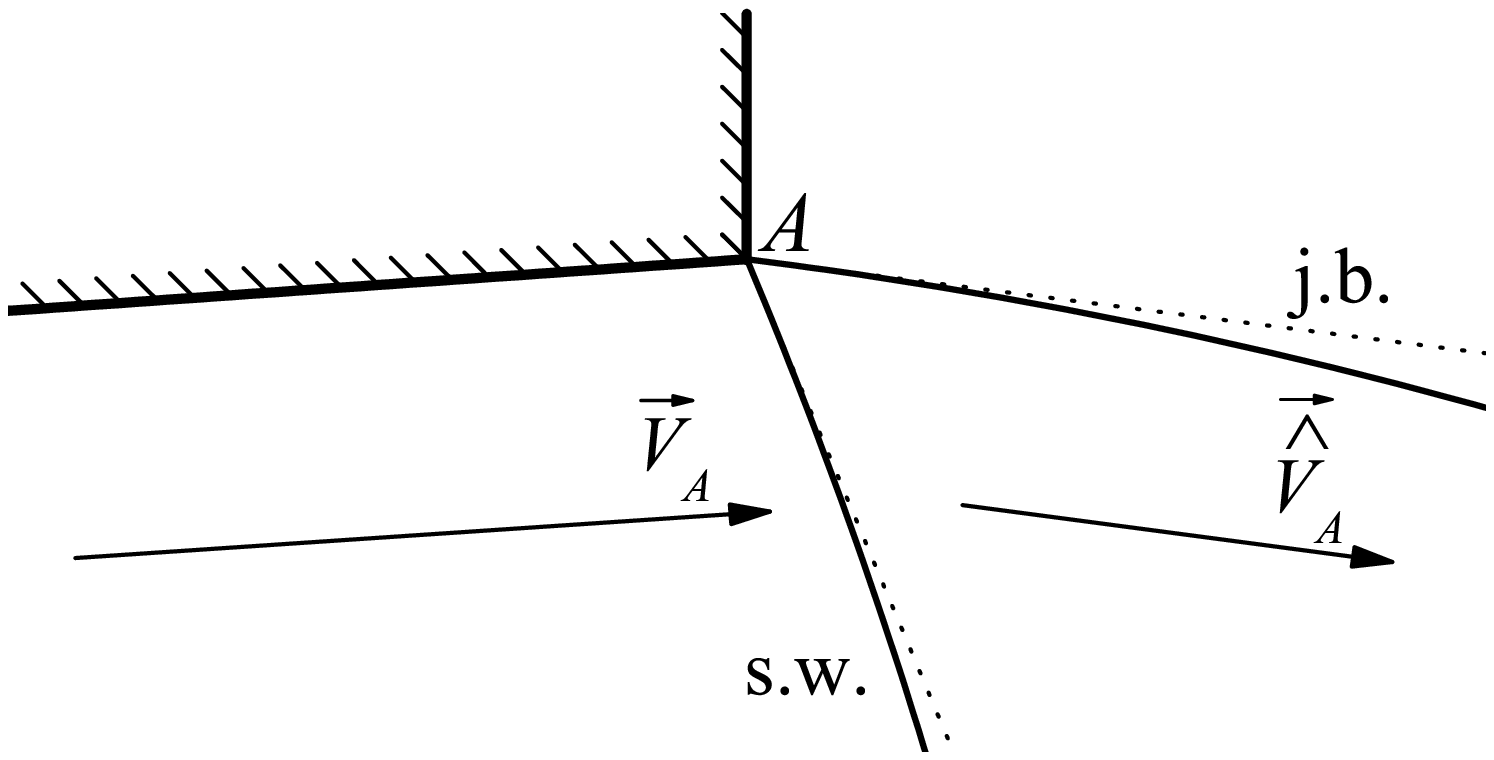}
\end{center}

\vspace{-0.3cm}

\caption{A free jet outflow from a nozzle in an overexpanded mode.}\label{SSW4A}

Solid lines~--- attached shock wave (s.w.) and jet boundary (j.b.). Dash lines~--- tangents to them.
$\vec V_A$, $\widehat V_A$~--- velocity vectors at the point $A$.
\end{figure}
\begin{figure}
\begin{center}
\mbox{\hspace{-0.5cm} \includegraphics[width=0.55\textwidth]{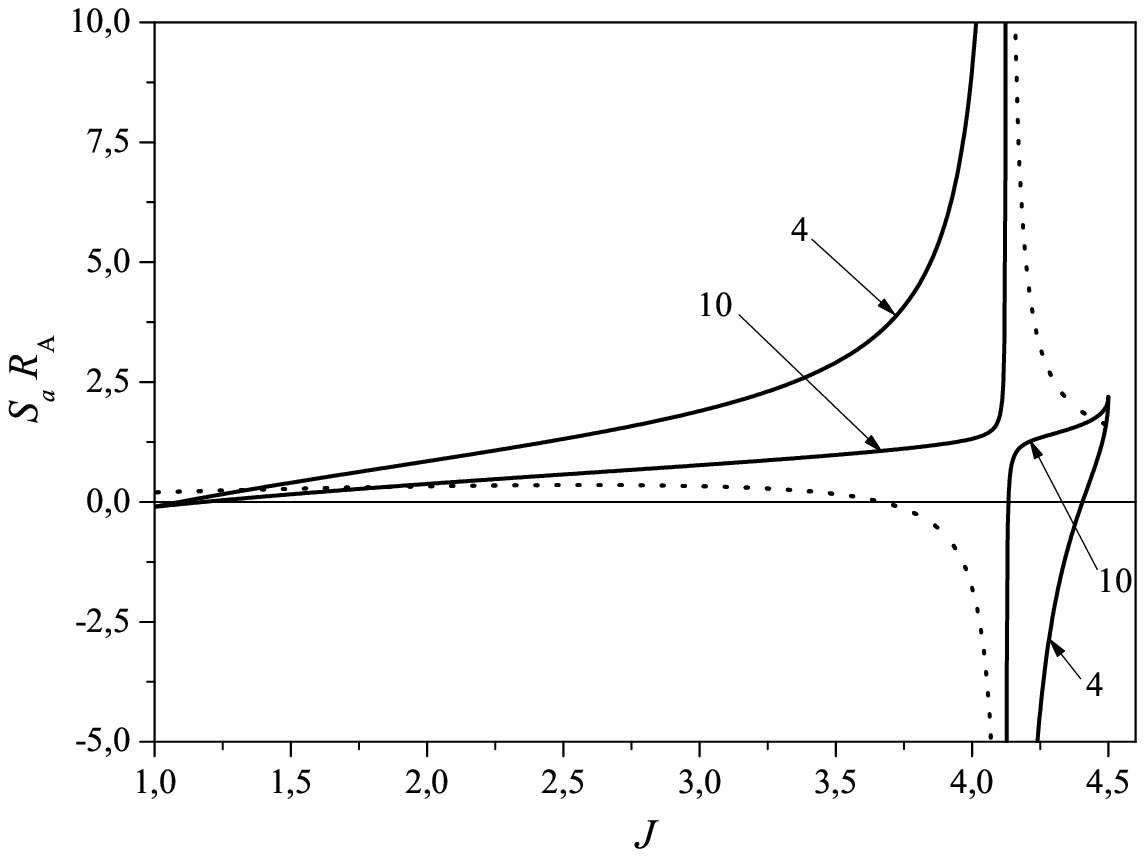} \hspace{-0.9cm} \raisebox{-0.9mm}{\includegraphics[width=0.55\textwidth]{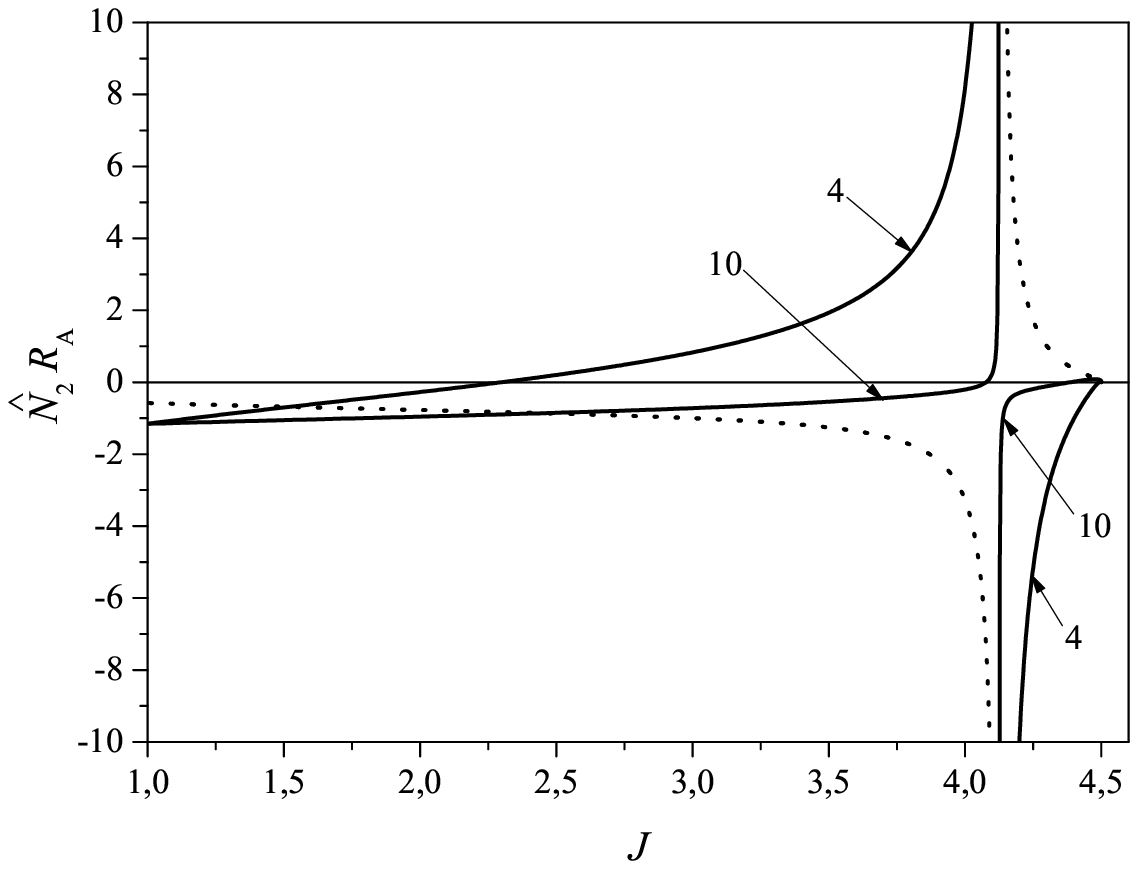}} \hspace{-0.5cm}}
\begin{picture}(0,0)
\put(-190,138){\large $(b)$}
\put(-384,138){\large $(a)$}
\end{picture}

\vspace{-0.3cm}

\caption{The dependences of the dimensionless curvatures of the shock $S_aR_A$ and the jet boundary $\widehat N_2 R_A$ on the edge of the nozzle on the shock intensity. Mach number $M_A=2$; plane flow~--- dotted line; axisymmetric flow with the nozzle opening half--angles $\varTheta_A=4^{\rm o}$ and $\varTheta_A=10^{\rm o}$~--- solid lines.}\label{SSW4}
\end{center}
\end{figure}

Let us consider a free jet outflow from a nozzle in an overexpanded mode (figure~\ref{SSW4A}). An attached shock wave forms on the nozzle lip $A$ and propagates downstrean in the jet. Let us determine the curvature of this shock $S_a$ in the point of its formation, and the curvature of the jet boundary. The nonisobaric factor $\widehat N_1$ is a linear function of $S_a$. Assuming the shock curvature $S_a$ to be $0$ or $1$, and using the specified upstream flow basic unevennesses, let us calculate the flow unevennesses behind the shock in both cases, using the algorithm twice. The resulting unevennesses are denoted $\widehat N_1^{(k)}$, $\widehat N_2^{(k)}$, $\widehat N_3^{(k)}$, $\widehat N_7^{(k)}$, here $k=0, \, 1$. Since $\widehat N_1=\widehat N_1^{(0)}(1-S_a)+\widehat N_1^{(1)}S_a=0$, we find $S_a=\widehat N_1^{(0)}/\left(\widehat N_1^{(0)}-\widehat N_1^{(1)}\right)$, then the basic unevennesses of the flow in the jet are determined.

Figure~\ref{SSW4}$(a)$ shows the curvature $S_a$ of the shock coming down from the nozzle edge, multiplied by $R_A$, versus the shock intensity $J$ for the Mach number $M_A=2$. Here $R_A$ is the distance from the edge of the nozzle to the center of the source. Both plane and axisymmetric flows are shown. In plane flows the curvature $S_a$ is independent of the nozzle opening half--angle $\varTheta_A$. The behaviour of the dependence in the plane flow ($\delta=0$) is qualitatively given in the paper (Uskov \& Chernyshov 2006).

Similar dependences for the curvature of the jet boundary (the curvature of the streamline, coming down from the edge of the nozzle) $\widehat N_2 R_A$ are shown in figure~\ref{SSW4}$(b)$. Figures~\ref{SSW4} show that for $M_A=2$ such value $J=4,125$ exists that in its vicinity the shock and the boundary curvatures grow infinitively. The existence of such $J$ was reported in (Uskov \& Chernyshov 2006).

\section{The isolines of gasdynamic parameters in the vicinity of a shock wave}\nopagebreak

In order to describe the gas flow in the vicinity of a shock wave let us trace the gasdynamic parameters isolines, i.e. curves with a constant value of a gasdynamic parameter $f$. In other words, the derivative of the parameter $f$ along the isoline turns to zero:\pagebreak[0]
\begin{equation}
\frac{\upartial f}{\upartial \xi}=\cos \alpha_f \frac{\upartial f}{\upartial \ell}+\chi \sin \alpha_f \frac{\upartial f}{\upartial n}=0,\label{1.XC}
\end{equation}
here $\vec \xi$ is the direction of the tangent to the isoline at the considered point, $\alpha_f$ is the isoline inclination angle with respect to the streamline.

In this section we consider only strong shocks, all parameters (except the rest enthalpy $h_0$) are discontinuous on the shock, and the isolines break. Let us define the angles $\alpha_f$, under which the isolines of various parameters come to the shock surface from the upstream flow, and the angles $\widehat \alpha_f$, under which they go out from the shock surface into the downstream flow. On the shock surface the derivatives in (\ref{1.XC}) are assumed to be one--sided. The angle $\alpha_f$ is measured from the direction $\vec \ell$ to $\vec \xi$ and is considered positive if the rotation is in the same direction as the rotation of $\vec \ell$ to $\vec \tau$. The range of the values is $\alpha_f \in \left(\sigma-\upi; \, \sigma\right]$ for the flow upstream the shock and $\widehat \alpha_f \in \left(\sigma-\beta-\upi; \, \sigma-\beta \right]$ for the flow downstream the shock wave. Below only the angles $\widehat \alpha_f$, under which the isolines of various parameters go into the flow downstream the shock, are studied. For brevity, the tilde character for angles $\alpha_f$ is below omitted.

The condition (\ref{1.XC}) uniquely determines the angle $\alpha_f$ value in the mentioned range, except for the case $\upartial f / \upartial \ell=\upartial f / \upartial n=0$; in the latter case $f$ is constant in the flow field in the vicinity of the point and the isolines degenerate.

Let us consider the isolines of entropy and the gas flow rest parameters. The gas flow rest parameters (pressure $p_0$, density $\rho_0$, temperature $T_0$, enthalpy $h_0$), and the gas entropy $s$ equal to its rest entropy $s_0$, remain constant along the streamlines outside of the shock surface (Chernyi 1994). Consequently, for these parameters either the streamlines are isolines and the equalities $\alpha_{p_0}=\alpha_{\rho_0}=\alpha_{T_0}=\alpha_{h_0}=\alpha_s=0$ hold, or the isolines degenerate.

In (M\"older 1979) thermodynamically perfect gas flows with constant rest enthalpy $h_0$ are considered. It is shown that under these assumptions the gas temperature, enthalpy, the speed of sound and the Mach number remain constant along the velocity magnitude isoline (isotach). Let us generalize this result for the flows of thermally perfect, calorically imperfect gas with constant rest enthalpy in the upstream flow. Really, since $h_0$ is constant, the equality $N_7 = 0$ holds according to (\ref{1.I}); then (\ref{Aij}) and (\ref{hN7}) give $\widehat N_7=0$ and therefore, the gas total enthalpy $h_0=\const$ in the whole flow field. The second formula (\ref{1.5}) shows that in this case the gas enthalpy $h$ is constant along the isotach. Since in the thermally perfect gas its enthalpy, temperature and the speed of sound are uniquely interrelated (appendix~\ref{App}), their isolines and the Mach number $M$ isolines coincide with the isotachs.

\begin{figure}
\vspace{-0.2cm}
\begin{center}
\includegraphics[width=\textwidth]{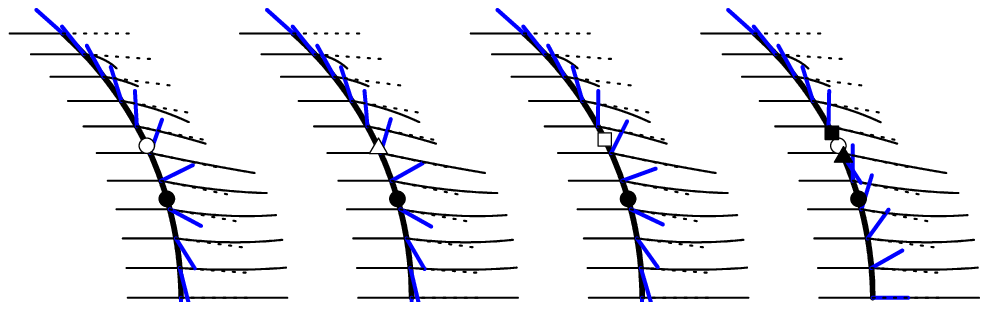}
\end{center}
\begin{picture} (0,0)
\put (210,129) {\large $(c)$}
\put (290,129) {\large $(d)$}
\put (50,129) {\large $(a)$}
\put (130,129) {\large $(b)$}
\end{picture}
\vspace{-1.2cm}
\begin{center}
\includegraphics[width=\textwidth]{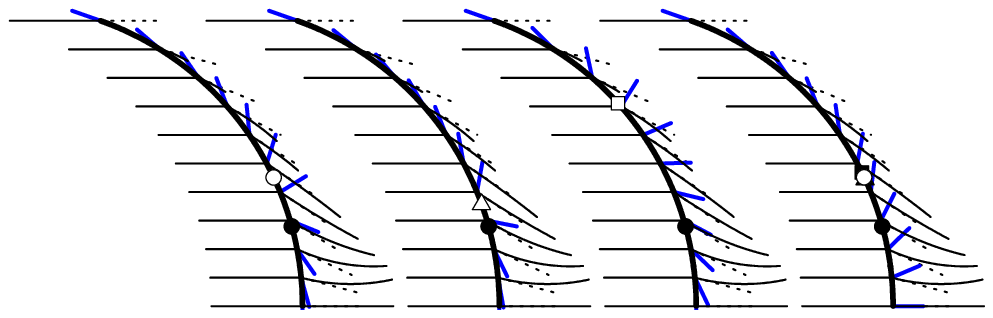}
\end{center}
\begin{picture} (0,0)
\put (198,135) {\large $(g)$}
\put (260,135) {\large $(h)$}
\put (58,135) {\large $(e)$}
\put (130,135) {\large $(f)$}
\end{picture}

\vspace{-0.9cm}

\caption {The flow patterns in the vicinity of a curved shock wave in a plane uniform initial flow with the Mach numbers $M = 1,5$ ($a$--$d$) and $M=3,0$ ($e$--$h$).} \label{isoline}

{\footnotesize solid lines~--- streamlines (the gas flows from left to right), dotted lines~--- tangents to them on the shock surface, short lines~--- isolines of various gasdynamic parameters
($a$, $e$: isobars, $b$, $f$: isopycnics, $c$, $g$: isotachs, $d$, $h$: isoclines);

open circle point~--- the Crocco point,

closed circle point ---~the constant pressure point (the Thomas point),

open square point~--- the point at which the isotach is perpendicular to the streamline,

closed square point~--- sonic point,

open triangular point~--- the point at which the isopycnic is perpendicular to the streamline,

closed triangular point~--- the point at which the flow in the shock wave turns to the maximum possible angle.}

\end{figure}

Following (M\"older 1979), let us take isobars ($f=p$), isopycnics ($f=\rho$), isotachs (\mbox{$f=V$}) and isoclines ($f=\varTheta$) into consideration. The algorithm given in \S~\ref{3} allows to determine numerically the derivatives with respect to the natural directions, and therefore find the streamline downstream the shock curvature $\widehat N_2$ and the isolines inclination angles $\alpha_f$ from (\ref{1.XC}). As an example, the flows of a diatomic thermodynamically perfect gas with the specific heats ratio $\gamma=1,4$ in the vicinity of a concave shock in a uniform plane upstream flow ($N_1=N_2=N_3=N_7=0$) with the Mach numbers $M=1,5$ and $M=3,0$ are considered. The shock curvature $S_a$ is assumed to be constant. The calculation results in the form of flow patterns are presented in figures~\ref{isoline}. The shock inclination angle $\sigma$ increases from top to bottom from a minimum value $\arcsin (1/M)$ to $\upi/2$. The shock intensity value increases, accordingly, from $J=1$ (degenerate shock) to a maximum value. The streamlines are shown with solid black lines, the dotted lines are the tangents to them at the points on the shock surface. The open circle point shows the position of a Crocco point; the streamline outcoming from this point into the downstream flow is non--curved (M\"older 1979; Uskov 1983; Adrianov \etal 1995): $\widehat N_2=0$. Above the Crocco point (in case $J <J_{\rm Crocco}$) the gas velocity inclination angle relative to the upstream flow increases downstream, below the Crocco point this angle decreases.

The short lines in figures~\ref{isoline} show the isolines directions in the flow downstream the shock, in dependence on the shock inclination angle $\sigma$. For the shock close to a degenerate one, i.e. at \mbox{$J \to 1$} (the upper edge of the picture), all isolines lie on the shock surface. For a normal shock wave, i.e., in case $\sigma \to\upi/2$ (the bottom edge of the picture), the isobar, the isopycnic, and the isotach also lie on the shock surface, whereas the isocline coincides with the straight streamline.

The closed circular point in the figures shows the constant pressure point (the Thomas point): $\widehat N_1=0$. In this point the gas pressure, density and velocity isolines coincide with the streamline, and the velocity vector polar angle isoline is perpendicular to it. The shock intensity at the constant pressure point $J_p$ for plane flows was calculated in (Adrianov \etal 1995). For arbitrarily curved shocks (including shocks in axisymmetric gas flows) M\"older \etal (2011) call this point the Thomas point or the generalized constant pressure point, keeping the index $p$ for its parameters. Above the Thomas point ($J<J_p$) the gas flow compressed in the shock wave continues to compress and to decelerate; below the Thomas point the flow downstream the shock is rarefied and accelerated. At the Crocco point the isocline lies on the streamline, and the isobar is normal to it. The closed square markes the sonic point; the downstream flow at this point has the velocity magnitude equal to the local speed of sound, $\widehat M=1$. At this point the isocline is perpendicular to the streamline; this result can be proved analytically from the relations (\ref{1.7}). At the point where the flow turns to the maximum possible angle, the isocline lies on the shock surface; the corresponding point is designated with a closed triangle. The latter two feature points and the Crocco point are clearly seen in figure~\ref{isoline}($d$). For larger Mach number values, including $M=3,0$ in figure~\ref{isoline}($h$), these three points merge and, consequently, the isocline inclination angle changes rapidly under small shock inclination angle variation.

\begin{figure}
\vspace{-0.5cm}
\begin{center}
\mbox{\hspace{-0.6cm} \includegraphics[height=0.465\textheight]{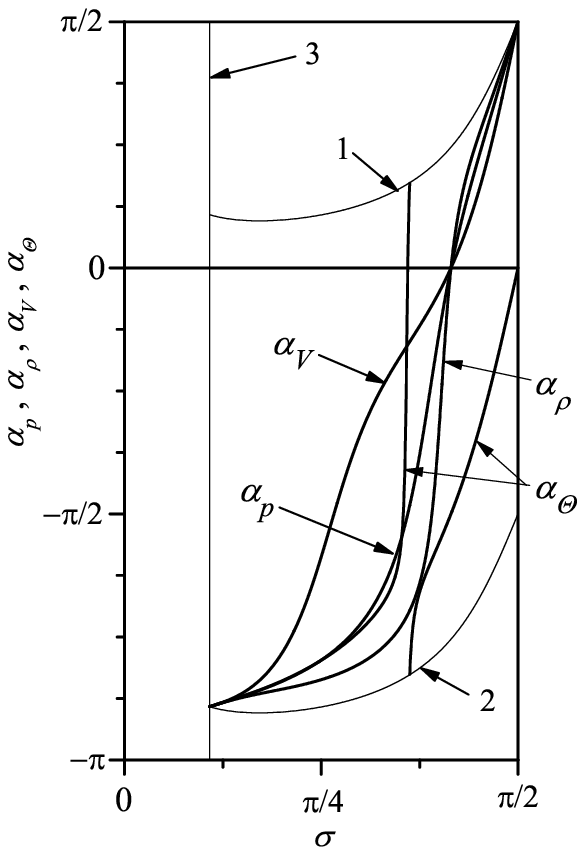} \hspace{-0.6cm}\includegraphics[height=0.465\textheight]{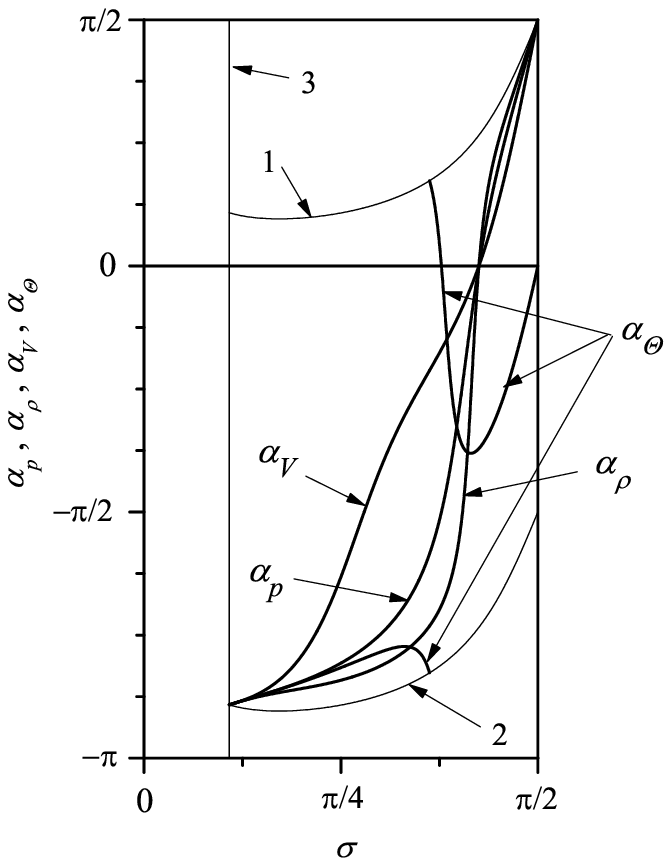}}\hspace{-0.4cm}\\
\begin{picture}(0,0)
\put(-192,254){\Large $(a)$}
\put(-10,254){\Large $(b)$}
\end{picture}
\end{center}

\vspace{-0.9cm}

\caption{Dependences of the gas pressure $\alpha_p$, velocity magnitude $\alpha_V$, density $\alpha_\rho$ and the velocity vector polar angle $\alpha_\varTheta$ isolines inclination angles downstream the shock wave on the shock inclination angle $\sigma$ for a uniform upstream flow with the Mach number $M=3,0$.}\label{Risun1}

{\footnotesize $a$~--- plane flow; $b$~--- axisymmetric flow with $\varTheta=0$, $N_4=\chi S_a$.}\vspace{0.7cm}

\vskip-25pt

\end{figure}

\begin{figure}
\vspace{-0.5cm}
\begin{center}
\mbox{\hspace{-0.7cm} \includegraphics[height=0.475\textheight]{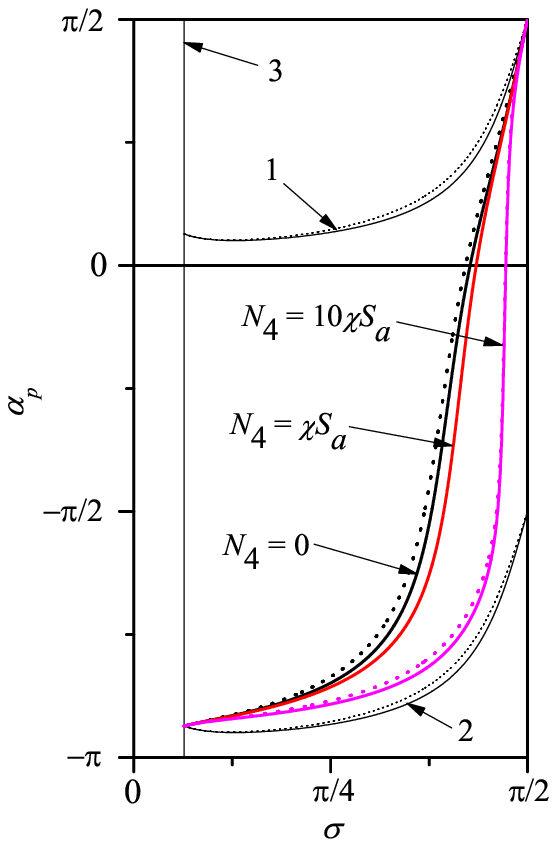} \includegraphics[height=0.475\textheight]{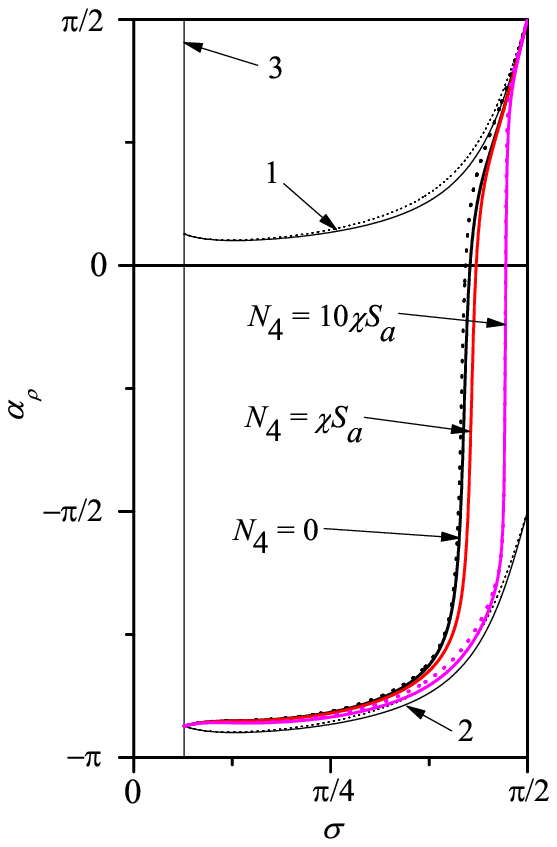} \hspace{-0.5cm}}\\
\vspace{-0.5cm}
\mbox{\hspace{-0.7cm} \includegraphics[height=0.475\textheight]{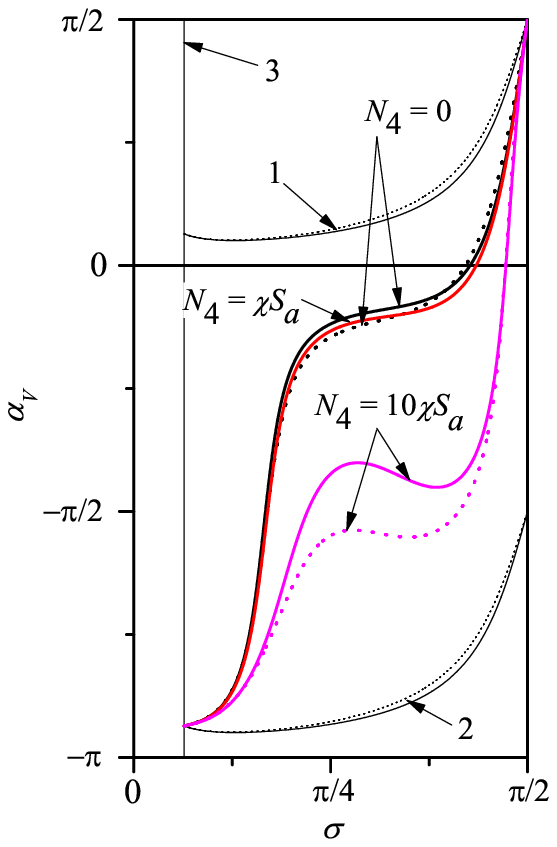} \includegraphics[height=0.475\textheight]{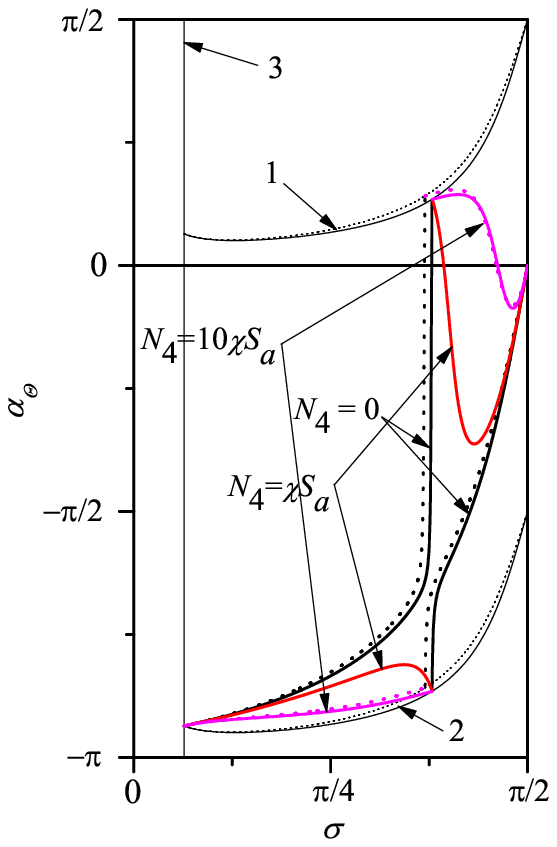} \hspace{-0.5cm}}\\
\begin{picture}(0,0)
\put(-190,259){\Large $(c)$}
\put(10,259){\Large $(d)$}
\put(-190,534){\Large $(a)$}
\put(10,534){\Large $(b)$}
\end{picture}
\end{center}
\caption{Dependences of the isolines inclination angles on the shock inclination angle $\sigma$ in a uniform plane flow ($N_4=0$)~--- black lines; in an axisymmetric flow for the curvature $N_4$ values $N_4=\chi S_a$ (red lines) and $N_4=10\chi S_a$ (magenta lines). Upstream flow Mach number $M=5,0$. Dotted lines~--- perfect gas, solid lines~--- model of oxygen for $p=10^5$~Pa, $T=300$~K.}\label{iso2}
\end{figure}

Figure~\ref{Risun1}$(a)$ shows the dependences of the angles $\alpha_p$, $\alpha_\rho$, $\alpha_V$ and $\alpha_\varTheta$ on the shock inclination angle $\sigma$ for plane flows of a diatomic thermodynamically perfect gas with the Mach number of a uniform upstream flow $M=3,0$ and the specific heats ratio $\gamma=1,4$. The dependences $\alpha_p(\sigma)$ and $\alpha_V(\sigma)$ coincide with the corresponding data of (M\"older 1979), so that they are represented on the graph by the same curves. These curves were used in the construction of the flow patterns in the vicinity of the shock wave in figure~\ref{isoline}($e$--$h$).

Let us determine the isolines inclination angles for a uniform axisymmetric flow (\mbox{$\varTheta=0$}). Using the formulae (\ref{1.7}), (\ref{Aij}) and (\ref{1.XC}) it can be shown that these angles are determined by the ratio $\chi N_4/S_a$ only. Figure~\ref{Risun1}$(b)$ shows these angles for a thermodynamically perfect gas flow with a Mach number $M=3,0$ for the relation between the shock curvatures $N_4=\chi S_a$.

The calculation showed that at low initial flow velocity ($M\leq 3$) use of the calorically imperfect gas model for oxygen and nitrogen does not lead to any significant deviations in the isolines inclination angles from the diatomic perfect gas model.

Figure~\ref{iso2} represents the same isolines inclination angles for a larger Mach number value $M=5,0$ for a diatomic thermodynamically perfect gas and oxygen in the previously described model. The uniform upstream flow parallel to the axis of symmetry at different distances from this axis is considered, in which the curvature $N_4$ takes the values \mbox{$N_4=\chi S_a$} and $N_4=10\chi S_a$. Plane flows are also shown for comparison ($N_4=0$). At this Mach number value, taking of oxygen calorical imperfection into account leads to a significant amendment to the model of a perfect gas.

The curves 1 and 2 in figures~\ref{Risun1}, \ref{iso2} show the angles between the flow velocity behind the shock $\widehat {\vec V}$ and the shock surface $(\sigma-\beta)$ and $(\sigma-\beta-\upi)$ and describe the boundaries of the region of the angles $\alpha_f$ definition. Line 3 in these figures corresponds to the \mbox{$\sigma=\arcsin (1/M)$}~--- the smallest possible angle $\sigma$ value for a given Mach number.

The vicinity of the sonic point on the shock wave surface is of special interest. In this point the gas velocity downstream a shock is equal to the local speed of sound (the Mach number $\widehat M=1$). In several works, including (Adrianov \etal 1995), the shock intensity and inclination angle at this point are determined as a function of the upstream flow Mach number: $J=J^*(M)$, $\sigma=\sigma^*(M)$. The parameters related to the sonic point are denoted by~$^*$.

In the flow of a thermally perfect gas with constant rest enthalpy the Mach number along the isotachs remains constant, therefore, the isotach passing through the sonic point is the sonic line and divides the flow region into the supersonic ($\widehat M>1$) and subsonic ($\widehat M<1$) ``patches''. In (M\"older \etal 2011) a classification of flows in the vicinity of the sonic point is proposed. This classification is based on the relative position of the outcoming streamline and sonic line.

The type of flow in the vicinity of the sonic point is determined by the following two criteria:

1) a) subsonic flow is accelerated to supersonic, \quad or

$ \, \, \, \, \, $ b) supersonic flow is decelerated to subsonic;

2) a) acoustic characteristics emerges from the sonic point in the supersonic flow (at the right angle to the flow line), \quad or

$ \, \, \, \, \, $ b) this characteristic degenerates into a point.

\begin{figure}
\begin{center}
\mbox{\hspace{-0.8cm} \includegraphics[height=5.8cm]{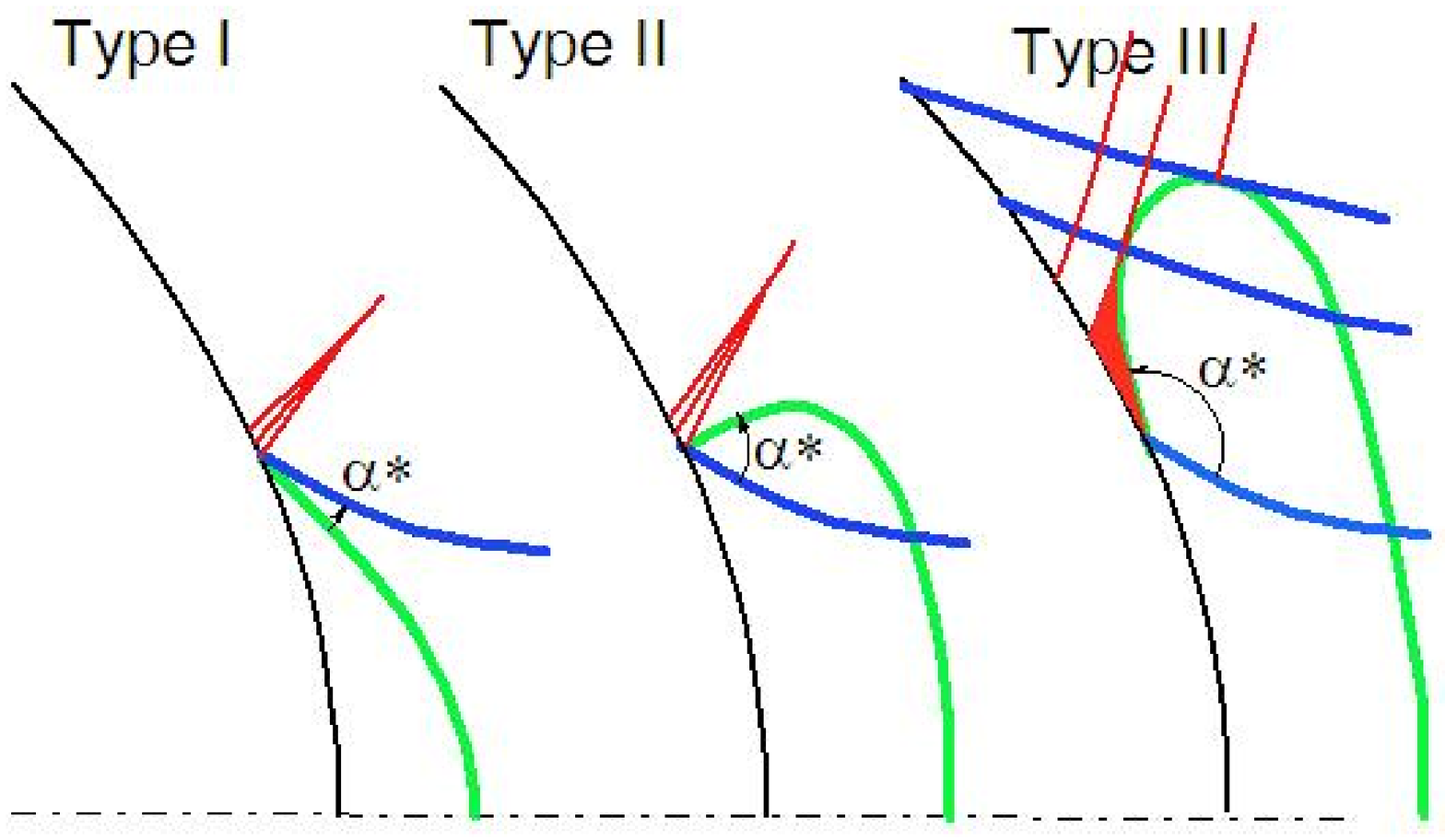} \hspace{-0.4cm} \raisebox{-0.5mm}{\includegraphics[height=5.8cm]{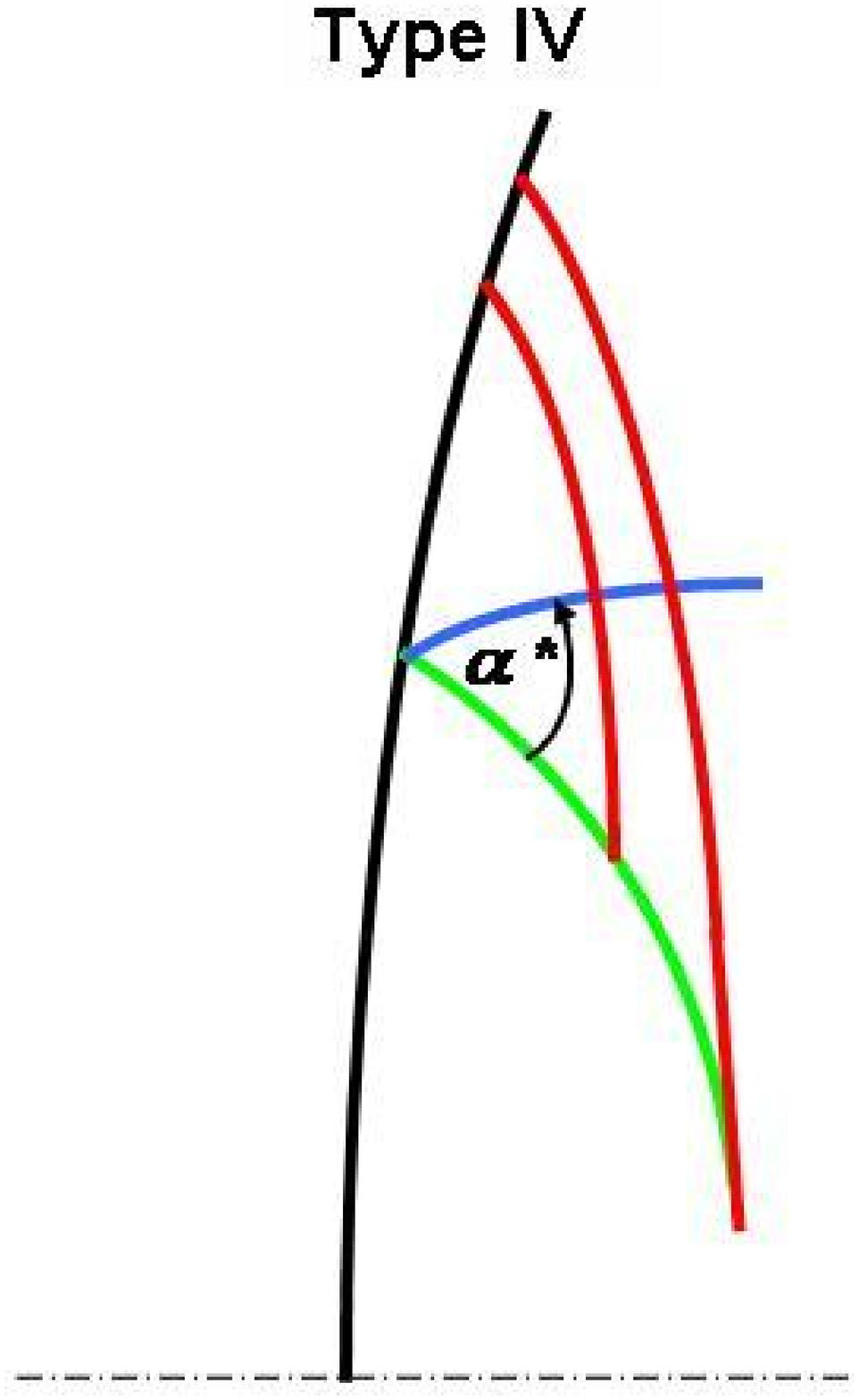}} \hspace{-0.4cm}}\\
\begin{picture}(0,0)
\put(-85,63){\large f.l.}
\put(-5,66){\large f.l.}
\put(85,66){\large f.l.}
\put(180,102){\large f.l.}
\put(-100,35){\large s.l.}
\put(-8,35){\large s.l.}
\put(85,35){\large s.l.}
\put(152,57){\large s.l.}
\put(-117,116){\large a.c.}
\put(-38,127){\large a.c.}
\put(50,159){\large a.c.}
\put(165,129){\large a.c.}
\put(-146,35){\large s.s.}
\put(-65,35){\large s.s.}
\put(25,35){\large s.s.}
\put(114,35){\large s.s.}
\end{picture}
\end{center}

\vspace{-0.6cm}
\caption{The schemes of four types of flows in the vicinity of the sonic point on the shock wave surface. The schemes of the types I, II \& III are borrowed from the paper (M\"older \etal 2011).}\label{Molder}

\footnotesize The upstream gas flow flows from left to right; the horizontal dash--dotted line at the bottom of the figure --- the axis of symmetry; the black line --- the shock wave surface (s.s.), the blue line --- the streamline (f.l.), the green line --- the sonic line (s.l.), the red lines --- the acoustic characteristics outgoing from the shock surface (a.c.).

\end{figure}

The criteria for each type of flow are listed in the first and the second columns of table~\ref{tabl2}. The flow types I, II and III were introduced in (M\"older \etal 2011) for shocks with two positive curvatures, i.e. facing its concavity to the upstream flow. For the shocks with the negative curvature in the flow plane (meridional half--plane) the flow type IV is possible; this type is introduced in the present paper. The schemes of the four types of flow are shown in figure~\ref{Molder}.
\begin{table}
\begin{center}
\begin{tabular}{ccccc}
The flow type&Criteria&&$\upartial \widehat M/\upartial \tau<0$&$\upartial \widehat M/\upartial \tau>0$\\
\hline
I&1a, 2a&&$0<\alpha_V^*<\upi/2$&$\alpha_V^*<-\upi/2$\\
II&1b, 2a&&$-\upi/2<\alpha_V^*<0$&---\\
III&1b, 2b&&$\alpha_V^*<-\upi/2$&$0<\alpha_V^*<\upi/2$\\
IV&1a, 2b&\quad&---&$-\upi/2<\alpha_V^*<0$\\
\end{tabular}
\end{center}
\caption{Types of flow in the vicinity of the sonic point on a shock wave}\label{tabl2}
\end{table}

The stated above criteria for the flow types can be mathematically formulated as follows:

1) a) in case $\upartial \widehat M/\upartial \tau<0$, $\alpha_V^*>0$ and in case $\upartial \widehat M/\upartial \tau>0$, $\alpha_V^*<0$, \quad or

$\, \, \, \, \,$ b) in case $\upartial \widehat M/\upartial \tau<0$, $\alpha_V^*<0$ and in case $\upartial \widehat M/\upartial \tau>0$, $\alpha_V^*>0$;

2) a) in case $\upartial \widehat M/\upartial \tau<0$, $\alpha_V^*>-\upi/2$ and in case $\upartial \widehat M/\upartial \tau>0$, $\alpha_V^*<-\upi/2$, \quad or

$\, \, \, \, \,$ b) in case $\upartial \widehat M/\upartial \tau<0$, $\alpha_V^*<-\upi/2$ and in case $\upartial \widehat M/\upartial \tau>0$, $\alpha_V^*>-\upi/2$.

\noindent Table~\ref{tabl2} allows to determine the flow type in accordance with the stated criteria in case the $\upartial \widehat M/\upartial \tau$ sign and the value $\alpha_V^*$ are prescribed (let us note that in (M\"older \etal 2011) the sign of the angle $\alpha^*$ is selected the opposite to the sign of $\alpha_V^*$ in this paper).

In case of a uniform upstream flow the velocity polar angle $\varTheta=\const$. The relations (\ref{1.6p}) and (\ref{1.XI}) relate the shock curvature $S_a$ with its inclination angle $\sigma$ derivative:
$$
S_a=\chi N_5=\chi \upartial \varOmega / \upartial \tau=\upartial \sigma / \upartial \tau.
$$
In the flow with a given Mach number the Mach number downstream the shock decreases if the shock inclination angle $\sigma$ increases (Chernyi 1994); therefore, the sign of $S_a$ is opposite to the sign of $\upartial \widehat M/\upartial \tau$ ($\upartial \widehat M/\upartial \tau<0$ on figures~\ref{F1}$a$ and~\ref{F1}$b$; $\upartial \widehat M/\upartial \tau>0$ on figures~\ref{F1}$c$ and~\ref{F1}$d$).

M\"older \etal (2011) show the isotach inclination angle $\alpha_V^*$ at the sonic point versus the upstream uniform flow Mach number $M$ for plane thermodynamically perfect gas flows ($S_b=0$) with $S_a>0$. They calculated $\alpha_V^*$ for a number of the ratio $S_a/S_b$ values for $S_a>0$, $S_b>0$ in axisymmetric flows. They determined the range of the parameters $M$ and $S_a/S_b$ values, corresponding to the three different types of flow. The results of (M\"older \etal 2011) show that at hypersonic speeds the isotach inclination angle $\alpha_V^*$ only slightly varies if the curvatures ratio $S_a/S_b$ changes. At smaller Mach number values this dependence is essential.

\begin{figure}
\vspace{-0.7cm}
\begin{center}
\includegraphics[width=\textwidth]{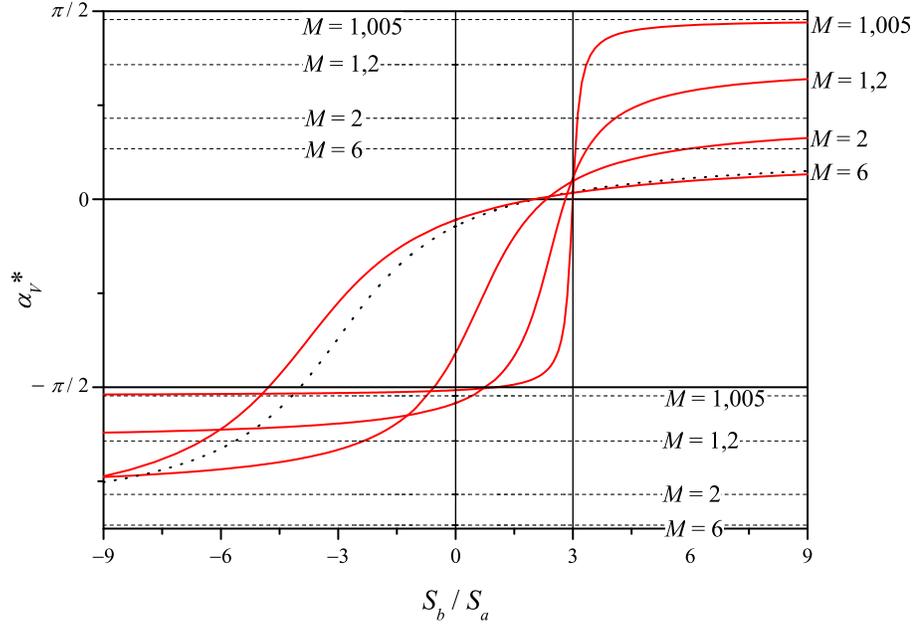}\\
\end{center}

\vspace{-0.6cm}
\caption{The dependence of the isotach inclination angle to the streamline in the sonic point on the shock wave $\alpha_V^*$ in a homogeneous axisymmetric flow on the curvatures ratio $S_b/S_a$ for different values of the upstream flow Mach number. The horizontal asymptotes of the curves are shown by dash lines. Black dotted lines~--- thermodynamically perfect gas, red solid lines~--- a model of oxygen for $p=10^5$~Pa, $T=300$~K (for $M\leq 2$ the dependences coincide).}\label{Risun2}

\end{figure}

The angle $\alpha_V^*$ vs. the curvatures ratio $S_b/S_a$ for different upstream flow Mach number $M$ for a uniform axisymmetric upstream flow is shown in figure~\ref{Risun2}. In case $S_b/S_a \geq 0$ the obtained results are consistent with the results presented in (M\"older \etal 2011) on figure~3$(a, \, b)$. The curves have horizontal asymptotes $(\sigma-\beta-\upi)$ in case $S_b/S_a \to -\infty$ and $(\sigma-\beta)$ in case $S_b/S_a \to\infty$. This result agrees with the fact that on a conical shock wave ($S_a=0$) the isolines coincide with the shock surface (M\"older 1979). If $M=1$, the stated dependence is a step function: the angle $\alpha_V^*$ is equal to $-\upi/2$ for $S_b/S_a<3$, $\upi/2$ for $S_b/S_a>3$ and not defined for $S_b/S_a=3$.

\begin{figure}
\vspace{-0.7cm}
\begin{center}
\includegraphics[width=\textwidth]{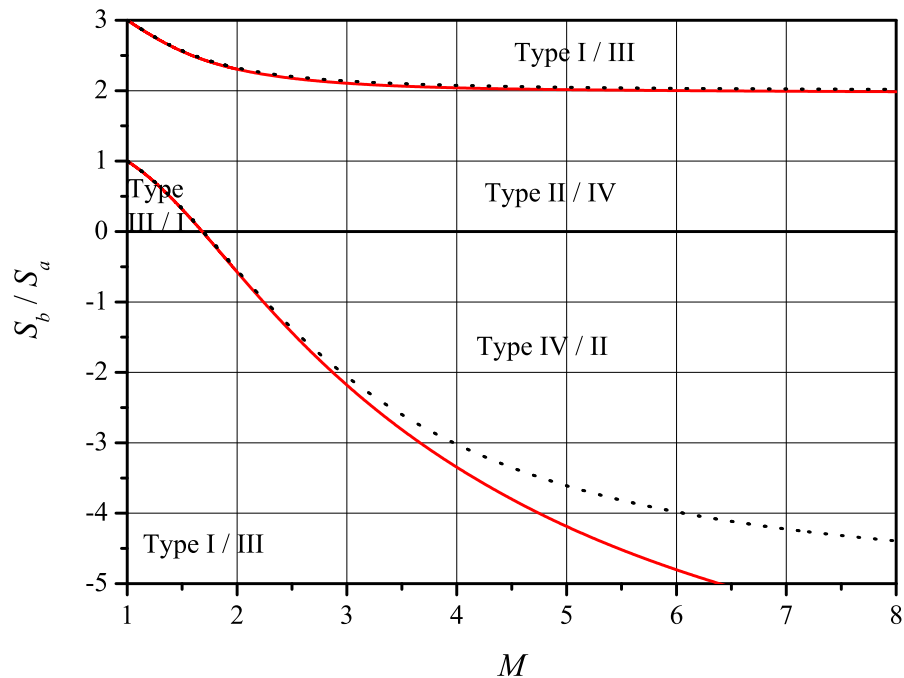}\\
\end{center}

\vspace{-0.8cm}
\caption{Types of flow in the vicinity of the sonic point on the shock wave in the plane $S_b / S_a$, $M$. Black dotted lines~--- thermodynamically perfect gas, red solid lines~--- a model of oxygen for $p=10^5$~Pa, $T=300$~K.}\label{Risun21}

\end{figure}

The transitional states between different types of flow are the situations the sonic line downstream the shock falls on the streamline ($\alpha_V^*=0$) and the sonic line falls on the acoustic characteristics ($\alpha_V^*=-\upi/2$). It can be shown that the curvatures ratios $S_b/S_a$ for the angles $\alpha_V^*=0$ and $\alpha_V^*=-\upi/2$ are given by the following formulae:
\begin{equation}
\frac{S_b}{S_a}=\frac{\rho  (\widehat{u}_\nu(u_\nu-\widehat{u}_\nu)+2\widehat \rho \widehat{h}_\rho)}{\widehat \rho^2 \widehat{h}_\rho\sin^2 (\sigma-\beta)}-\frac{1}{\sin^2 (\sigma-\beta)} \qquad \qquad \mbox{for} \quad \alpha_V^*=-\upi/2; \label{1.R1}
\end{equation}
\begin{equation}
\frac{S_b}{S_a}=\frac{\rho}{\widehat \rho}-1+\frac{\rho  (2\widehat{u}_\nu(u_\nu-\widehat{u}_\nu)+3\widehat \rho \widehat{h}_\rho)}{\widehat \rho^2 \widehat{h}_\rho \sin^2 (\sigma-\beta)} \qquad \qquad \mbox{for} \quad \alpha_V^*=0. \label{1.R2}
\end{equation}
The right--hand sides in (\ref{1.R1}) and (\ref{1.R2}) depend on the upstream flow parameters only. For perfect gas flows, they are determined by its Mach number only; consequently, the type of flow in the vicinity of the sonic point on the shock is determined by $M$ and $S_b/S_a$. The domains of $S_b/S_a$ and $M$ variation corresponding to the four different types of flow are shown in figure~\ref{Risun21}. The upper numbers give the types of flow for the shocks incoming to the axis of symmetry ($S_b>0$); the lower numbers~--- for the shocks outgoing from the axis ($S_b<0$). For $S_a, \, S_b>0$ the figure is consistent with the results of (M\"older \etal 2011).

In the limit $M \to 1$ it follows from (\ref{1.R1}), (\ref{1.R2}) that the curvatures ratio $S_b / S_a \to 1$ and 3, respectively. This result is valid for a gas with arbitrary thermodynamic properties at the sonic point on the shock wave. For perfect gas it is numerically obtained in (M\"older \etal 2011).

\section{Conclusions}\nopagebreak

The first--order problem for plane and axisymmetric flows of non--viscous non--heat\-con\-duc\-tive gas in a state of thermodynamic equilibrium in the vicinity of a shock wave is solved. In other words, the space gasdynamic parameters derivatives in the flow on the downstream side of the shock wave surface are determined. For this purpose the differential dynamic compatibility conditions on shock waves for gas flows satisfying an arbitrary thermodynamical equation of state are used. A set of basic flow unevennesses is singled out: the flow nonisobaric factor along the streamline $N_1$, the streamline curvature $N_2$, the flow vorticity factor $N_3$, and the flow nonisoenthalpy factor $N_7$. The basic flow unevennesses are defined in terms of derivatives with respect to natural directions. The gasdynamic parameters derivatives in an arbitrary point of the flow are expressed through these unevennesses.

On the basis of the DDCC, the algorithm for the basic unevennesses of the flow downstream a shock wave calculation is proposed. In this algorithm, the basic unevennesses of the flow upstream the shock and its curvatures are supposed known. The influence coefficients of the shock curvatures and of the basic unevennesses of the initial flow on the basic flow unevennesses downstream the shock are calculated. The comparison of results in the thermodynamically perfect gas model and the calorically imperfect gas model of oxygen is presented. Using the same algorithm for the case of the overexpanded jet outflow from a nozzle the curvature of the shock, coming down from the edge of the nozzle, and the curvature of the jet boundary are obtained.

The gas flow in the vicinity of the shock wave is described by the isobars, isopycnics, isotachs and isoclines for plane and axisymmetric flows of a thermodynamically perfect gas and oxygen in the thermally perfect gas model. The angles under which the isolines go out into the flow downstream the shock are calculated. The position of the characteristic points on the shock surface are determined. In these points the isolines coincide either with the streamline, or with the perpendicular to it.

The behaviour of the isolines in the vicinity of the sonic point on the shock surface is studied. Four possible types of flow in this vicinity are identified. The types of flow in the $M$, $S_b/S_a$ plane for any curvatures signs are analytically determined. For shocks with two positive curvatures the types of flow were numerically established in (M\"older \etal 2011).

\begin{acknowledgments}
This research is financially supported by the St.--Petersburg State University (project No~6.50.1556.2013) and the Russian Foundation for Basic Research (project No~12--08--00826--a).
\end{acknowledgments}

\oneappendix \section{The derivatives of the thermodynamic functions}\label{App}

The dependence of the gas specific enthalpy $\tilde h=\tilde h(p, \, s)$ on its pressure $p$ and specific entropy $s$  is a form of the canonical equation of gas state, which gives the full information on its thermal and calorical properties (Sivukhin 1975; Chernyi 1994). The tilde sign is used for the enthalpy as a function of the variables $p$, $s$. The gas temperature and density are given by partial derivatives
\begin{equation}
\tilde h_p=\frac{1}{\rho}; \qquad \tilde h_s=T.\label{C.1}
\end{equation}
For the mixed derivative $\tilde h_{ps}$ we have
$$
\tilde h_{ps}=\frac{\upartial}{\upartial s}\left(\frac 1\rho\right)\biggl|_{p=\const}=\frac{\upartial T}{\upartial p}\biggl|_{s=\const}>0,
$$
since the adiabatic compression is always accompanied by temperature and pressure increase. Therefore, for a fixed $p$ the derivative $\tilde h_p$ as a function of $s$ increases, and a unique value of $s$ matches a fixed value of $\rho$. The specific entropy $s$ and enthalpy $h$ can, consequently, be numerically determined as functions of $p$ and $\rho$: $s(p, \, \rho)$, \mbox{$h(p, \, \rho)=\tilde h(p, \, s(p, \, \rho))$}.

For the differential of entropy we have
\begin{equation}
\mathrm{d}s=s_p\mathrm{d}p+s_\rho \mathrm{d}\rho,\label{C.6}
\end{equation}
and for the speed of sound we obtain
\begin{equation}
a^2=\frac{\upartial p}{\upartial \rho}\biggl|_{s=\const}=-\frac{s_\rho}{s_p}.\label{C.7}
\end{equation}

For the density, according to (\ref{C.1}), we have
$$
\frac{1}{\rho}=\frac{\upartial \tilde h}{\upartial p}\biggl|_{s=\const}=\frac{\upartial}{\upartial p}h(p, \, \rho(p, \, s))\biggl|_{s=\const}=h_p+h_\rho\cdot\frac{\upartial \rho}{\upartial p}\biggl|_{s=\const}=h_p+\frac{h_\rho}{a^2},
$$
wherefrom
\begin{equation}
a^2=\frac{\rho h_\rho}{1-\rho h_p}, \qquad h_p=\frac{h_\rho s_p}{s_\rho}+\frac{1}{\rho}.\label{C.8} \label{1.III}
\end{equation}

Writing the first formula (\ref{C.1}) in differentials, we obtain
\begin{equation}
\tilde h_{pp}\mathrm{d}p+\tilde h_{ps}\mathrm{d}s=\mathrm{d}\tilde h_p=-\frac{\mathrm{d}\rho}{\rho^2}.\label{1.C}
\end{equation}
Comparing (\ref{1.C}) with (\ref{C.6}), we obtain expressions for the derivatives of entropy and enthalpy:
\begin{equation}
s_p=-\frac{\tilde h_{pp}}{\tilde h_{ps}}>0; \qquad s_\rho=-\frac{1}{\rho^2\tilde h_{ps}}<0, \qquad h_\rho=\tilde h_s s_\rho=Ts_\rho<0.\label{C.3}
\end{equation}

Let the gas be thermally perfect, i.e. satisfying the Clapeyron equation of its state: $p=\rho RT$. In this case the gas enthalpy is the function of its temperature only: $h=h(T)$ (Sivukhin 1975); therefore
\begin{equation}
h_\rho=h_T\cdot\frac{\upartial T}{\upartial \rho}\biggl|_{p=\const}=-h_T\cdot \frac p{\rho^2 R}=-h_T\frac T\rho; \qquad h_p=h_T\cdot\frac{\upartial T}{\upartial p}\biggl|_{\rho=\const}=h_T\cdot \frac 1{\rho R},\label{B6}
\end{equation}
here $h_T=\mathrm{d}h/\mathrm{d}T$. Substitution of (\ref{B6}) in (\ref{C.8}) gives for the speed of sound
\begin{equation}
a=\left(\frac{h_TRT}{h_T-R}\right)^{1/2},\label{B7}
\end{equation}
which is also a function of the gas temperature only.

\end{document}